\documentclass[11pt]{article} % Change from {jss} to {article}

\newcommand{\pkg}[1]{{\fontseries{b}\selectfont #1}}
\newcommand{\proglang}[1]{\textsf{#1}}

\usepackage{fancyvrb}
\usepackage{color}

\newenvironment{CodeChunk}{ \vspace{0.5em} }{ \vspace{0.5em} }

\DefineVerbatimEnvironment{CodeInput}{Verbatim}{
  fontsize=\small,
  fontshape=sl, % Slightly slanted like JSS
  xleftmargin=2em
}

\DefineVerbatimEnvironment{CodeOutput}{Verbatim}{
  fontsize=\small,
  xleftmargin=2em
}
\newcommand{\code}[1]{\texttt{\detokenize{#1}\unskip}}

\DefineVerbatimEnvironment{Code}{Verbatim}{
  fontsize=\small,
  xleftmargin=2em,
  baselinestretch=1
}

\usepackage[margin=1in]{geometry} % Standard margins
\usepackage{mathrsfs}   
\usepackage{amssymb}
\usepackage{multirow}
\usepackage{booktabs}
\usepackage{threeparttable}
\usepackage{subcaption}
\usepackage[T1]{fontenc}
\usepackage{lmodern}
\usepackage{textcomp}
\usepackage{float}
\usepackage{hyperref} % For clickable links

\usepackage[authoryear,round]{natbib} % Adds (Author, Year) support
\usepackage{graphicx}
\usepackage{authblk}

\title{\pkg{CCMnet}: A Software Package for Network Generation with Congruence Class Models}

\author[1]{Ravi Goyal}
\author[1]{Victor De Gruttola}
\author[1]{Natasha K. Martin}
\author[2]{Lior Rennert}
\author[3]{Jukka-Pekka Onnela}

\affil[1]{Division of Infectious Diseases and Global Public Health, University of California, San Diego}
\affil[2]{Department of Public Health Sciences, Clemson University}
\affil[3]{Department of Biostatistics, Harvard University}

\date{} 

\begin{document}
\maketitle

\begin{abstract}
We introduce \pkg{CCMnet}, an \proglang{R} package designed to generate network ensembles that accurately reflect the uncertainty inherent in empirical data. While traditional network modeling often results in ensembles with fixed property values or model-determined levels of variability, \pkg{CCMnet} enables a continuous spectrum of variability for network properties, including edge counts, degree distribution, and mixing patterns. By defining probability distributions directly over congruence classes of networks, the package allows researchers to specify the uncertainty in network properties across the generated ensemble to match a specific sampling design or empirical distribution. Furthermore, this formulation provides a principled framework that encompasses several classic models (e.g., Erd\H{o}s--R\'{e}nyi model, stochastic block models, and certain exponential random graph models) that implicitly share this structural basis, while offering the flexibility to specify arbitrary, even non-parametric, distributions for network properties. \pkg{CCMnet} implements a Markov chain Monte Carlo (MCMC) framework to sample from these models. The utility of the package is illustrated by generating posterior predictive network ensembles representing school friendship networks.\end{abstract}


%% include your article here, just as usual
%% Note that you should use the \pkg{}, \proglang{} and \code{} commands.

\section{Introduction}\label{introduction}

Often research questions in network analysis require the ability to generate realistic network ensembles rather than a single representative graph \citep{leung2023simulating, hiram2022disease, zhu2025hepatitis}. Ideally, the network ensemble should appropriately reflect the estimated uncertainty in the network structure \citep{wang2014sample}; this uncertainty can arise from the nature of the sampling of the population of interest \citep{krivitsky2017inference} or from stochasticity in the generation process \citep{robins2007introduction}. In this paper, we present the \pkg{CCMnet} package that enables generating networks with investigator specified uncertainty in network properties within the Congruence Class Model framework.

The level of uncertainty associated with a particular network property derived from a network ensemble can be view as a spectrum, from zero to a high degree of uncertainty. For many probabilistic models, the uncertainty they generate for a given network property occupies a specific point along this spectrum. At one end of this spectrum, a hard constraint specifies that the distribution of a network property from repeated realizations is a point mass, that is, it has zero variability \citep{bollobas2011random}. In network models with variability in the structure, that variability is often fixed by model-imposed constraints; therefore, these models still reside at a single point on this spectrum. For example, exponential-family random graph models (ERGMs) define probability distributions on graphs by maximizing entropy \citep{lusher2013exponential}, which results in a fixed and model-imposed level of uncertainty determined by sufficient statistics \citep{newman2018networks}. This is analogous to a Poisson distribution where, although individual realizations vary, the variability across an ensemble of realizations is linked to the mean; once the mean is specified, the variance of the ensemble is fixed. In many empirical contexts, such as an epidemiologist modeling disease transmission within a school, a fixed variability may be inadequate if the model must reflect the specific margin of error or uncertainty inherent in the underlying survey data. 

\begin{figure}
\centering
\includegraphics[width=0.9\textwidth]{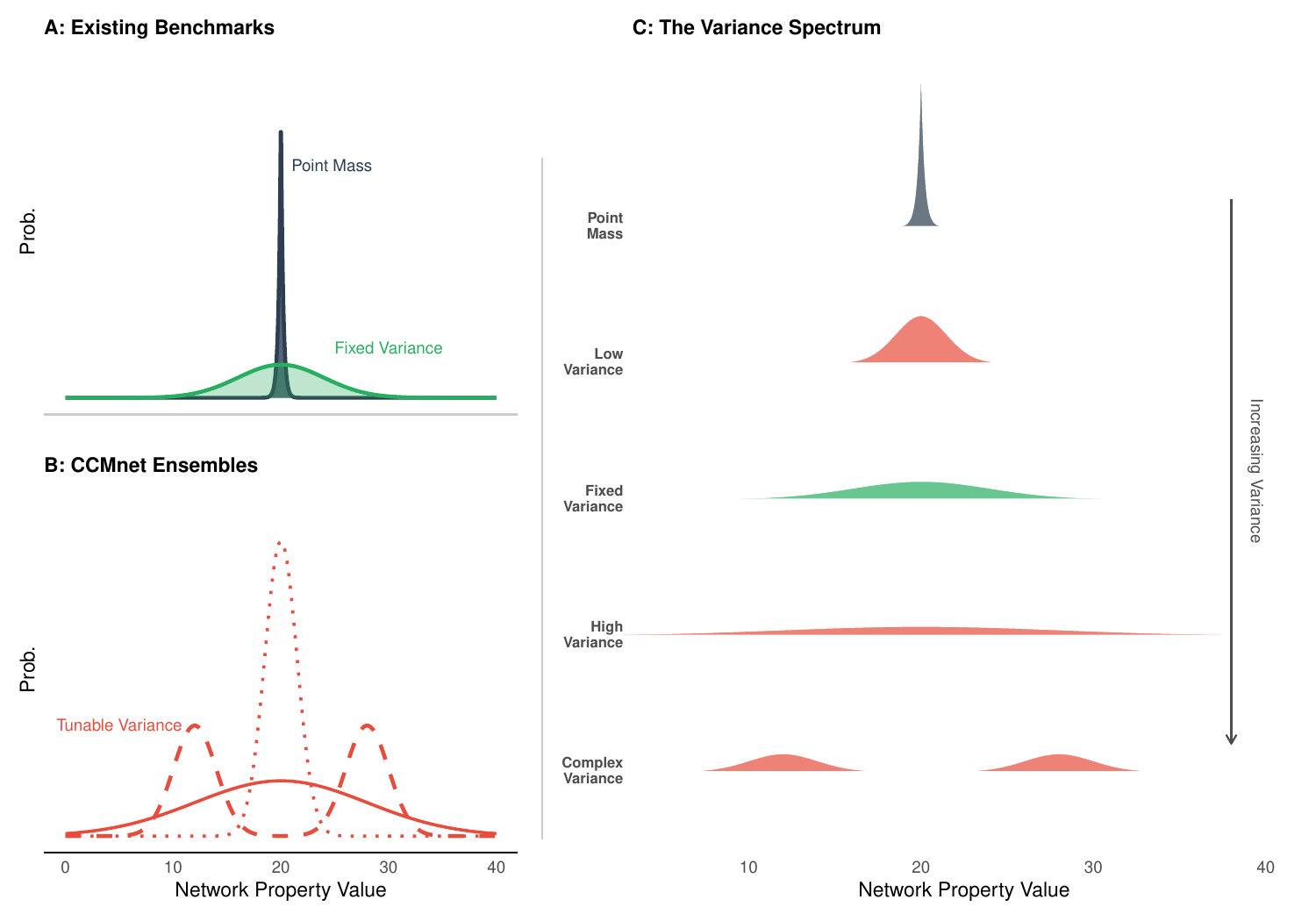}
\caption{Conceptual illustration of network uncertainty. All the plots show distributions for a network property, which can represent edge count, degree distribution, or mixing patterns. In Panel A, the two distributions represents variance under a hard-constraint (dark blue) and a soft-constraint where the variance is fixed (green). Panel B shows the flexibility of CCMs to model a broad range of uncertainty for the same network property (red curves). Panel C demonstrates the continuous spectrum of uncertainty by mapping the network property distributions in Panels A and B to where they reside on the spectrum.}
\label{fig:Conceptual_net_prop}
\end{figure}

Figure~\ref{fig:Conceptual_net_prop} (Panel A) provides an illustration of this limitation by displaying the distributions of a network property under different modeling regimes. When the model imposes a hard constraint on the property, the distribution is a point mass (dark blue distribution); that is, the values for all the network realizations are the same. In the models where a soft-constraint is imposed, the values of the property will vary across networks within the ensemble, but typically the model imposes a fixed level of uncertainty on the distribution (green distribution). Although these models provide valuable tools for network simulation, they do not offer a  mechanism for adjusting the degree of uncertainty in network properties to reflect the amount of information available in the empirical data. By allowing for a continuous spectrum of variability, researchers can tune the ensemble to more accurately reflect the structural uncertainty present in real-world observations.

Congruence Class Models (CCMs) were developed to address this shortcoming by providing a framework for generating network ensembles that explicitly propagate uncertainty in network properties estimated from data \citep{goyal2014sampling}. CCMs assign probability distributions to network properties; the concentration of these distributions reflects the amount of available information. When data are sparse or noisy, constraints are loose, yielding high-variance ensembles; as data become more informative, constraints tighten, yielding network ensembles that increasingly resemble a hard constraint. This perspective allows the network simulation to incorporate sampling variability and statistical uncertainty. Furthermore, CCMs provides a common framework for characterizing a number of existing network models as special or limiting cases. Figure~\ref{fig:Conceptual_net_prop} (Panel B) provides a conceptual depiction of how different levels and types of variability can be captured (red curves). Panel C maps the network property distributions in Panels A and B to where they reside on a conceptualized uncertainty spectrum.

In this paper, we introduce \pkg{CCMnet}, a software package for generating networks using CCMs. Section~\ref{sec:background} reviews some existing approaches to network generation, and shows how CCMs compare to models that impose hard- and soft- constraints on network properties. Section~\ref{sec:CCM} introduces the CCM framework, formally defines congruence classes, and shows how probability distributions on network properties induce distributions on graphs. Section~\ref{sec:software} describes the architecture of \pkg{CCMnet}. Section~\ref{sec:net_props} provides details on the network properties currently implemented, while Section~\ref{sec:prob_distr} details probability distributions on the properties implemented and the process to include additional distributions. 

Section~\ref{sec:syntax} focuses on syntax for the main network sampling function and presents supporting diagnostic functions. Section~\ref{sec:illustrations} provides illustrative examples of generating networks across different network properties and probability distributions on the properties. Section~\ref{sec:application} provides a practical application for researchers in fields like epidemiology and public health by generating posterior predictive network ensembles representing school friendship networks. This example demonstrates how \pkg{CCMnet} allows investigators to propagate the uncertainty induced by specific sampling designs into the resulting network structures. Finally, Sections~\ref{sec:add_applications} and~\ref{sec:discussion} discuss additional applications and directions for future methodological development.

\section{Constraints on Network Properties} \label{sec:background}

Network generation is a broad area of research, but for the purposes of contextualizing CCMs, it is useful to focus on statistical or probabilistic models, rather than mechanistic network models. Statistical models differ from mechanistic models in that they define probability distributions over networks rather than simulating networks from generative processes or behavioral rules \citep{mattie2025review}. Importantly, network properties associated with statistical models can also be classified according to whether their value is exactly the same for all realizations (hard constraints) or vary across realizations (soft constraints).

\subsection{Hard-Constraints on Network Properties}

A hard constraint specifies that all network realizations will have an identical value for a given property, even if other properties or network structure vary. Two canonical examples are the number of edges in the $G(n,m)$ model \citep{erdds1959random} and the degree sequence in the configuration model \citep{blitzstein2011sequential, fosdick2018configuring}. In the $G(n,m)$ model, a network of $n$ nodes is sampled uniformly from the set of all graphs that contains exactly $m$ edges. This ensures the total edge count is identical across all realizations, resulting in a variance of zero for that property. Several software packages implement this, such as the \code{sample_gnm} function in the \pkg{igraph} package \citep{igraph_rcran}. Similarly, the configuration model preserves each node's degree exactly; \pkg{igraph} provides the \code{sample_degseq} function for this purpose \citep{igraph_rcran}.

\subsection{Soft-Constraints on Network Properties}

Soft constraints define the distribution of a network property in expectation rather than exactly. The Erd\H{o}s--R\'{e}nyi--Gilbert $G(n,p)$ random graph model \citep{gilbert1959random}, wherein each edge is present independently of others with probability $p$, is the simplest example of a model with soft constraint \citep{bollobas2011random}. The $G(n,p)$ model allows for control of the expected number of edges while allowing variability in individual network realizations; specifically, the number of edges follows a binomial distribution, which is in contrast to the $G(n,m)$ model where the number is fixed. There are several \proglang{R} software packages available to simulate from the $G(n,p)$ model, including, \pkg{igraph} (\code{sample_gnp}) \citep{igraph_rcran}. Stochastic Block Models (SBMs) extend this idea by specifying probabilities of connection among predefined groups, allowing the number of edges within and between groups to vary probabilistically based on a binomial distribution \citep{holland1983stochastic}; see the \proglang{R} packages \pkg{sbm}, \pkg{blockmodels}, and \pkg{igraph} for examples of implementation of this model \citep{sbm_rcran, blockmodels_rcran, igraph_rcran}. ERGMs generalize further, allowing flexible parametric specification of network statistics such as edge counts, degree distributions, or triadic closure, to be specified in expectation \citep{robins2007introduction}; the variance of these statistics follows from entropy maximization \citep{lusher2013exponential}. The \pkg{statnet} package in \proglang{R} is the primary software to fit and simulate from ERGMs \citep{HunterHandcock2008}. To generate networks from ERGMs, the user first defines the probability of the network ensemble through a vector of coefficients ($\theta$) that correspond to various network motifs. Once the coefficients are specified, generating networks can be done using the \code{simulate} function in the \pkg{ergm} package \citep{HunterHandcock2008}.

\subsection{Spectrum of Constraints on Network Properties}

\begin{figure}
\centering
\includegraphics[width=0.9\textwidth]{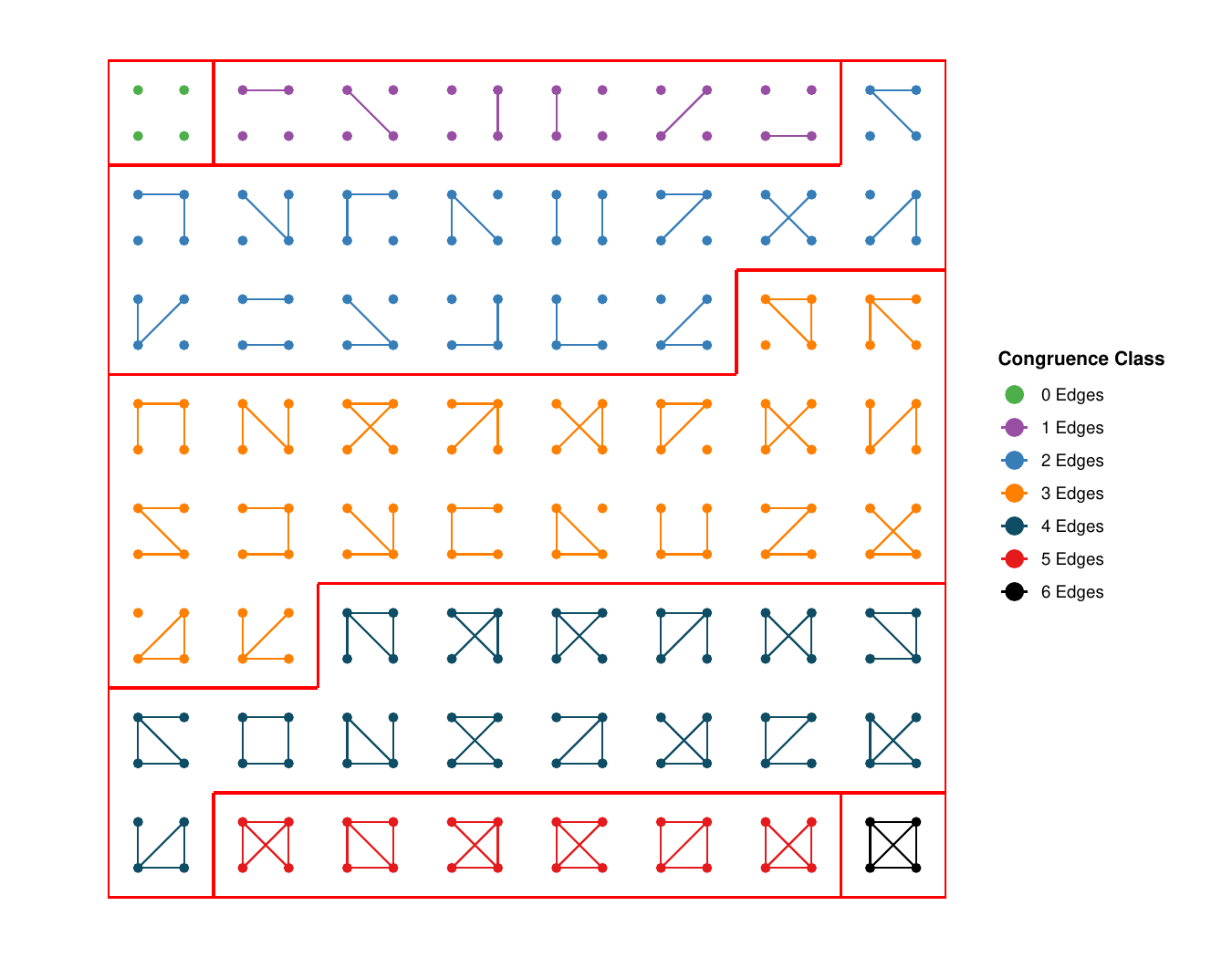}
\caption{The space of all $2^{{4 \choose 2}} = 64$ undirected binary graphs of size $n=4$, partitioned into congruence classes based on edge count. Each cell represents a unique graph, with colors and red boundaries indicating the seven distinct congruence classes (0 to 6 edges).}
\label{fig:Conceptual_congruence_class}
\end{figure}

The network models reviewed above implicitly partition the space of all possible networks into congruence classes defined by properties such as edge counts, degree sequences, or mixing matrices. Each model induces a probability distribution over these classes, while treating individual networks within the same class as equally likely. Figure~\ref{fig:Conceptual_congruence_class} provides an illustration of the idea by presenting all undirected binary graphs of size equal to $4$. The networks are colored-coded on the basis of their number of edges; hence, the colors correspond to unique congruence classes defined by edge count. This visualization demonstrates the structural grouping underlying many network models, including the $G(n,m)$ model, $G(n,p)$ model, and ERGMs with a single term for the number of edges. For example, the $G(n,m)$ model would assign probability $0$ to classes associated with graphs that have edges not equal to $m$ and one to the class with networks with $m$ edges, each graph within this class has probability ${n \choose m}^{-1}$. By defining user-specified probability distributions directly on the property-based classes themselves, it is possible to have a spectrum of constraints on a network property.

\section{Congruence Class Models}\label{sec:CCM}

CCMs formalize the perspective of equivalence classes, which is an implicit framework of many existing network models. Rather than beginning with a probability distribution on individual graphs, CCMs start by specifying a probability distribution over congruence classes \citep{goyal2014sampling}. A probability distribution on individual networks is then induced by assigning equal probability to all networks within each class. This inversion makes explicit the probabilistic assumptions that are only implicit in many network models and provides a unifying framework encompassing both hard- and soft-constraint approaches. By specifying distributions at the class level, CCMs can reproduce the behavior of classical network models such as $G(n,p)$, configuration models, SBMs, and ERGMs with discrete covariates, while also allowing for flexible modeling of the probability distribution on congruence classes. In the following subsection, we formalize the CCM framework.

\subsection{Congruence Classes of Networks}

In this section, we follow the established notation and terminology used to define CCMs \citep{goyal2024investigating}. Let $\mathscr{G}_n$ denote the set of all undirected binary networks with $n$ nodes. Let $\phi$ denote an algebraic mapping from $\mathscr{G}_n$ to network summary statistics (e.g., degree distribution and mixing patterns). Let $c_{\phi}(x) := \{g : \phi(g) = x, g \in \mathscr{G}_n\}$ denote the inverse image associated with $\phi$. These inverse images partition the network space $\mathscr{G}_n$ into congruence classes according to the chosen summary statistics. Let $U_{\phi} = \{x \colon g \in \mathscr{G}_n, \phi(g)=x\}$ denote the set of all possible property values. Finally, let $\vert c_{\phi}(x) \vert$ denote the cardinality of the congruence class $c_{\phi}(x)$.

\subsection{Congruence Class Models}

A CCM specifies a probability distribution $P_{\phi}(x \vert \theta)$ over congruence classes indexed by $U_{\phi}$. Because the model assigns uniform probability to all networks within the same class, the probability of observing a particular network is

\begin{equation} \label{eq:networkprob_static}
P_{\mathscr{G}_n}(g \vert \theta) =\left(\frac{1}{\vert c_{\phi}(\phi(g)) \vert} \right) P_{\phi}(\phi(g) \vert \theta).
\end{equation}

\noindent This construction ensures that networks within the same class are equally likely, while allowing flexible specification of class-level probabilities to capture diverse network-generating mechanisms \citep{goyal2023framework}. The choice of $P_{\phi}(x \vert \theta)$ can be based on empirical data and either be parametric (e.g., Poisson distribution), or nonparametric. This flexibility, which is central to the \pkg{CCMnet} implementation, allows researchers to incorporate empirical uncertainty directly into the network generation process.

\subsection{Simple Examples}

To demonstrate the flexibility of the CCM framework, Table~\ref{table:supp_edges_updated} presents three ways to assign probabilities to the congruence classes of all $64$ undirected graphs with $n=4$, which are visualized by color in Figure~\ref{fig:Conceptual_congruence_class}. Following the logic in \citet{goyal2023estimating}, we use this small-scale case to show how different class-level distributions, $P_{\phi}(x \mid \theta)$, influence the probability of drawing any specific network, $P_{\mathscr{G}_n}(g \mid \theta)$. The congruence classes are defined by edge count (Column 1), which is directly mapped to the colors in Figure~\ref{fig:Conceptual_congruence_class} (Column 2). The size of each class is shown in Column 3; the sum of this column is $64$, which is the total number of graphs of size $n=4$. The probability of drawing any individual network ($P_{\mathscr{G}_n}(g)$) within a class is simply the class-level probability ($P_{\phi}(x)$) divided by the class size ($|c_{\phi}(x)|$); for example, the values in Column 5 are calculated as Column 4 divided by Column 3. The Uniform CCM (Columns 4--5) assigns equal probability to each of the seven edge-count classes ($1/7 \approx 0.143$). In contrast, the Binomial CCM (Columns 6--7) reproduces the behavior of a $G(n,p)$ model where $p=0.5$, meaning every one of the 64 individual networks is equally likely ($1/64 \approx 0.0156$). Finally, the Non-traditional CCM (Columns 8--9) is a probability mass assignment that does not follow traditional parametric form. This direct control over congruence class probabilities is a core feature of both CCM and the \pkg{CCMnet} software.

\begin{table}[H]
\caption{Illustration of CCMs for the number of edges ($n=4$). Each row represents a congruence class defined by a specific edge count (Column 1), mapping directly to the colors in Figure~\ref{fig:Conceptual_congruence_class} (Column 2). The size of each class is shown in Column 3. Columns 4--9 compare three distinct ways to assign probability mass ($P_{\phi}$) and the resulting individual network probabilities ($P_{\mathscr{G}_n}$).}
\centering
\small
\begin{tabular}{lcc|cc|cc|cc}
\toprule
\textbf{Edges} & \textbf{Class Color} & \textbf{Size} & \multicolumn{2}{c}{\textbf{Uniform CCM}} & \multicolumn{2}{c}{\textbf{Binomial CCM}} & \multicolumn{2}{c}{\textbf{Non-traditional CCM}} \\
($x$) & (Fig~\ref{fig:Conceptual_congruence_class}) & $|c_{\phi}(x)|$ & $P_{\phi}(x)$ & $P_{\mathscr{G}_n}(g)$ & $P_{\phi}(x)$ & $P_{\mathscr{G}_n}(g)$ & $P_{\phi}(x)$ & $P_{\mathscr{G}_n}(g)$ \\
\midrule
0 & Green      & 1  & 0.143 & 0.1430 & 0.016 & 0.0156 & 0.050 & 0.0500 \\
1 & Purple     & 6  & 0.143 & 0.0238 & 0.094 & 0.0156 & 0.200 & 0.0333 \\
2 & Blue       & 15 & 0.143 & 0.0095 & 0.234 & 0.0156 & 0.100 & 0.0067 \\
3 & Orange     & 20 & 0.143 & 0.0071 & 0.313 & 0.0156 & 0.350 & 0.0175 \\
4 & Dark Blue  & 15 & 0.143 & 0.0095 & 0.234 & 0.0156 & 0.150 & 0.0100 \\
5 & Red        & 6  & 0.143 & 0.0238 & 0.094 & 0.0156 & 0.100 & 0.0167 \\
6 & Black      & 1  & 0.143 & 0.1430 & 0.016 & 0.0156 & 0.050 & 0.0500 \\
\bottomrule
\end{tabular}
\label{table:supp_edges_updated}
\end{table}

\section[CCMnet]{\pkg{CCMnet} Software Architecture} \label{sec:software}

\begin{figure}[H]
\centering
\includegraphics[width=400pt]{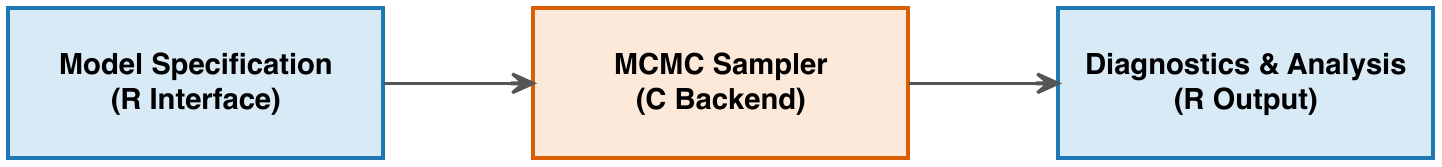}
\caption{The modular architecture of \pkg{CCMnet}. The model is specified via the \proglang{R} interface, where input validation occurs. The Markov chain Monte Carlo (MCMC) algorithm is implemented in the \proglang{C} backend. Resulting network and MCMC diagnostics are then returned to \proglang{R} for analysis.}
\label{fig:Conceptual_architecture}
\end{figure}

The \pkg{CCMnet} package is designed with an architecture that separates model input and output in \proglang{R} from intensive computations in \proglang{C}. This separation ensures a user-friendly interface while maintaining the performance required for network simulation. Figure~\ref{fig:Conceptual_architecture} provides a schematic overview of this flow.

\subsection{Interface and Input Validation}

The \proglang{R} layer serves as the primary user interface to call the network generation function, \code{sample_ccm}; syntax for the function is provided in Section~\ref{sec:syntax}. Its main role is to facilitate model specification and ensure input integrity before the Markov chain Monte Carlo (MCMC) algorithm is executed. The core specification requires the following primary inputs: the network properties being modeled, the target probability distributions and associated parameters, and the MCMC parameters. Issues with inputs trigger a useful error message. Once validated, the model configuration and initial network state are passed to the compiled \proglang{C} backend. There are four categories of validation performed; descriptions of the categories are provided below:

\begin{itemize}
\item \textbf{Network Statistics and Properties:} The function validates that the requested network properties are among those currently implemented. If a covariate-based property is selected, the interface confirms that the covariate pattern is provided, contains no missing values, and matches the specified population size (which is also verified to be a positive integer).

\item \textbf{Probability Distribution Parameters:} A rule-based system checks that the parameters are numeric and meet the mathematical requirements of the chosen distribution. For example, it verifies that rate and scale parameters are strictly positive for a Gamma distribution.

\item \textbf{Dimensionality and Compatibility:} The interface ensures the number of properties matches the number of distributions and parameter sets. It also performs structural checks to ensure parameter dimensions match the requirements of the network property, such as verifying that a mixing matrix is square and its dimensions correspond to the number of unique groups specified.

\item \textbf{MCMC Parameters:} The interface verifies that the sample size, burn-in period, and interval between samples are all positive integers.
\end{itemize}

\subsection{Computational Backend}

The core of \pkg{CCMnet} is a Metropolis-Hastings MCMC procedure written in \proglang{C}. The MCMC samples networks from $\mathscr{G}_n$ to generate an ensemble consistent with the target distribution $P_\phi(x|\theta)$.

The MCMC chain is initialized at a starting network $g^{(0)}$. By default, the algorithm begins with a random graph (Erdős-Rényi) with an initial density of 0.05. If this density results in a network that violates hard constraints---such as exceeding a maximum degree specified in a degree distribution target---the density is iteratively lowered until a valid initial graph is realized. Alternatively, users may also provide a specific starting network. 

For each MCMC iteration $t$, the following five steps are conducted:

\begin{enumerate}
\item Proposal: To create a proposal network $gp^t$ for the Metropolis-Hastings MCMC algorithm, an edge is selected using the Tie-No-Tie (TnT) toggle algorithm, which more efficiently explores $\mathscr{G}_n$ compared to selecting edges to toggle at random \citep{morris2008specification}. TnT chooses to either toggle an existing edge or a non-edge in the current network $g^t$ with equal probability, yielding a proposal network $gp^t$ that differs from $g^t$ by exactly one dyad. Let $q(gp^t|g^t)$ and $q(gp^t|g^t)$ denote the probability of proposing $gp^t$ given $g^t$ and vice versa, respectively.   

\item Network Statistics: The MCMC calculates the network summary statistics for the current network ($x^t = \phi(g^t)$) and proposal ($xp^t = \phi(gp^t)$). To efficiently calculate these statistics, CCMnet utilizes  internal structures and  logic developed for the \pkg{statnet} suite \citep{Hunter_2008_statnet}. 

\item Congruence Class Ratio: Sampling from a CCM requires evalation of the ratio of the cardinalities of the congruence classes: 

\begin{equation} \label{eq:congruence_class_ratio}
    \frac{\vert c_\phi(xp^t) \vert}{\vert c_\phi(x^t)\vert}. 
\end{equation}

\noindent The implementation of this calculation is specific to each network property. The combinatorial formulas to estimate these ratios are provided in \citet{goyal2014sampling}. 

\item Target Density: The target probability density for the $g^t$ $P_\phi(x^t|\theta)$ and $gp^t$ $P_\phi(xp^t|\theta)$ is evaluated for the current and proposal statistics $x^t$ and $xp^t$, respectively. \pkg{CCMnet} supports a range of distributions, including the Poisson, Gamma, and Dirichlet-Multinomial; details of the implemented distributions as well as how to include additional ones can be found in Section 5.3.

\item Acceptance/Rejection: The proposal is accepted with probability:$$\min \left( 1, \frac{P_\phi(xp^t|\theta)}{P_\phi(x^t|\theta)} \times \frac{|c_\phi(x^t)|}{|c_\phi(xp^t)|} \times \frac{q(g^t|gp^t)}{q(gp^t|g^t)} \right)$$If accepted, $g^{(t+1)} = gp^t$; otherwise, $g^{(t+1)} = g^t$.
\end{enumerate}

\subsection{Output and Diagnostics}

Upon completion of the MCMC chain, the resulting network and a diagnostic matrix of MCMC summary network statistics are returned to \proglang{R} and encapsulated in a \code{"sample_ccm"} object. The network within this object is represented as an igraph object, allowing users to transition from network generation to network analysis using the tools available in the \pkg{igraph} package \citep{igraph_rcran}. Each row of the MCMC sample matrix represents the summary statistics of the selected network property modeled; the number of rows is governed by user specified \code{sample_size} MCMC parameter. Furthermore, the returned object is designed to work with standard \proglang{R} methods for visualization and assessment. Users can generate trace plots and density estimates to monitor sampler convergence and verify that the simulated network properties align with the target distribution; additional details are provided below. 

\section{Network Properties Implemented} \label{sec:net_props}

The network properties implemented in \pkg{CCMnet} are motivated by the $dk$-series framework, which defines increasingly constrained random graph ensembles  \citep{mahadevan2006systematic, orsini2015quantifying}; additional details of the $dk$-series framework are provide below. \pkg{CCMnet} extends this framework in two important ways. First, rather than holding their values fixed, tunable soft constraints are placed on the network properties. These constraints are governed by user-specified probability distributions, allowing for a flexible representation of structural uncertainty. Second, \pkg{CCMnet} supports network properties that depend on node-level covariates. This extension is particularly significant for empirical research wherein network structure is often driven by node-level attributes, such as homophily or social stratification \citep{mcpherson2001birds}.

\subsection{Topological-based Network Properties}

The $dk$-series framework defines a systematic hierarchy of random graph ensembles. \pkg{CCMnet} implements the $0k$-, $1k$-, $2k$-, and $2.1k$-distributions, which correspond, respectively, to the following network properties: the number of edges, degree distribution, joint degree distribution, and the joint degree distribution together with average clustering \citep{mahadevan2006systematic, orsini2015quantifying}. Below we provide details of these properties:

\begin{itemize}
    \item \textbf{Edges (0k):} The total number of edges within the graph.
    
    \item \textbf{Degree distribution (1k):} A vector where the $j^{th}$ entry represents the number of nodes having degree $j$.
    
    \item \textbf{Joint degree distribution (2k):} Represented as a degree mixing matrix, where the $(i,j)$ entry represents the number of edges between nodes having degree $i$ and degree $j$, capturing degree-degree correlations (degree assortativity).
    
    \item \textbf{Joint degree distribution and clustering (2.1k):} Includes the joint degree distribution along with the total number of closed triads (triangles) in the network.
\end{itemize}

\subsection{Covariate-based Network Properties}

\pkg{CCMnet} incorporates node-level attributes by defining congruence classes based on categorical or discretized covariates. In this framework, network structure is summarized by the mixing patterns within and between covariate-defined groups. These covariate-based properties can be implemented independently or in combination with topological properties. The following covariate-based summarizes are currently implemented:

\begin{itemize}
    \item \textbf{Attribute mixing matrix:} A matrix where the $(i,j)$ entry represents the number of edges between nodes belonging to covariate groups $i$ and $j$.
    
    \item \textbf{Attribute mixing matrix and degree distribution:} A combination property that includes the attribute mixing matrix alongside separate degree distributions calculated for each distinct covariate group.
\end{itemize}

\section{Probability Target Distribution on Congruence Classes} \label{sec:prob_distr}

An important advantage of the CCM framework is that the inclusion of additional probability target distributions on congruence classes does not require any additional mathematical theory or a change in the underlying MCMC algorithm. Unlike adding a new network property, which requires the derivation of the congruence class cardinality ratio (Equation~\ref{eq:congruence_class_ratio}), new probability target distributions on congruence classes over existing properties can be added by including code for calculating the probability mass function and including details in an R configuration file. In this section, we provide specifications for existing target probability distributions as well as details on including additional distributions.

\subsection{Specification}

The target distribution, $P_\phi(x|\theta)$, on congruence classes is specified by providing the name of the distribution (e.g., gamma) and its associated parameters (e.g., scale and rate). In the \proglang{R} interface, the distribution names are provided in a vector and the associated parameters are contained in a list, where each element corresponds to a specific network property in the vector. Each of these parameter elements in the list is itself a list containing up to two vectors of parameters (referred to as $p1$ and $p2$) required by the chosen distribution.

For vector-valued network properties (such as degreedist or mixing), $p1$ and $p2$ are often vectors of the same length as the property. For example, to specify independent gamma distributions for a 10-bin degree distribution, $p1$ would be a numeric vector of $10$ shape parameters and $p2$ would be a numeric vector of 10 rate parameters. However, to two vectors can be of different lengths. For example, if the distribution is a multivariate normal distribution, the list entry would contain the mean vector ($p1$) and covariance matrix ($p2$) as the first and second entries, respectively. Table~\ref{tab:dist_specs} details the expected data structure for the currently implemented distributions.

\begin{table}[t!]
\caption{Parameter data structures for target distributions in \pkg{CCMnet}.}
\centering
\begin{tabular}{lll}
\toprule
Distribution & \code{p1} & \code{p2} \\
\midrule
Poisson & Vector of means ($\lambda$) & \code{NULL} \\
Gamma & Vector of shapes ($\alpha$) & Vector of rates ($\beta$) \\
Normal & Vector of means ($\mu$) & Vector of variances ($\sigma^2$) \\
Beta & Vector of Shape 1 ($\alpha$) & Vector of Shape 2 ($\beta$) \\
Dirichlet-Multinomial & Vector of concentration ($\alpha$) & \code{NULL} \\
Multivariate Normal & Mean vector ($\mu$) & Covariance matrix ($\Sigma$) \\ 
Uniform & \code{NULL} & \code{NULL} \\
Lognormal & Vector of log means & Vector of log standard deviations\\
Non-Parametric & Vector of probabilities & NULL \\
\bottomrule
\end{tabular}\label{tab:dist_specs}
\end{table}

\subsection{Extension}

The modularity of \pkg{CCMnet} allows for the incorporation of new target distributions without modifying the MCMC algorithm. To demonstrate this, we detail the two building blocks required to implement a new distribution using the Normal distribution as a concrete example.

\subsubsection{The R Configuration Layer} The first block is a structured entry in an \proglang{R} configuration file (\code{CCMnet_prob_distr_config.R}). This entry defines how the distribution interacts with the input validation checks and the target probability diagnostics. For the normal distribution, the configuration is specified as follows:

\begin{Code}
"normal" = list(sub_code = 1, 
                mean_bool = TRUE, 
                var_bool = TRUE, 
                use_solve_var = FALSE,
                rules = list(p1 = c("is_numeric"), 
                             p2 = c("is_numeric", 
                                    "all_positive", 
                                    "match_length")),
                sampler = function(p, n, ...) {
                   mu <- p[[1]]; sigma <- p[[2]]
                   matrix(rnorm(n * length(mu), 
                          mean = mu, 
                          sd = sqrt(sigma)), 
                          nrow = n, 
                          byrow = TRUE)
                },
                valid_network_prop = c("edges", "density", "degreedist", 
                                       "degmixing", "triangles", "mixing")
),
\end{Code}

This configuration contains four critical components that ensure modularity:

\begin{itemize}

\item \textbf{Identification (\code{sub\_code}):} A unique integer ID used by the \proglang{C} backend to select the corresponding probability distribution.

\item \textbf{Validation Rules (\code{rules}):} A list of constraints for the parameter entries \code{p1} and \code{p2}. For the Normal distribution, \code{p1} (means) must be numeric (\code{is_numeric}), while \code{p2} (variances) must be, in addition, strictly positive (\code{all_positive}), and match the vector length of \code{p1} (\code{match_length}). These rules are processed by the centralized validation framework described in Section 4.1.

\item \textbf{Target Probability Distribution Sampler (\code{sampler}):} A function that generates values for the network statistics from the target probability distribution for diagnostics. 

\item \textbf{Compatibility (\code{valid\_network\_prop}):} A whitelist of network properties for which this distribution is mathematically valid. This prevents users from inadvertently specifying incompatible models.

\end{itemize}

\subsubsection{The C Backend Probability Kernel:} The second block is the implementation of the log-probability density in a \proglang{C} backend file (\code{CCMnet_netprop_prob_dist.c}). The backend is designed to be property-agnostic; it simply receives a vector of statistics for $g^t$ (\code{g_stats}) and $gp^t$ (\code{gp_stats}) and vectors of target distribution parameters (\code{p1} and \code{p2}) to compute the log-density for $g^t$ (\code{g_pdf}) and $gp^t$ (\code{gp_pdf}). Note, all univariate probability distributions are coded as being a product of independent variables for greater usage across network properties. As only the probability ratio between the proposal and current network summary statistics needs to be calculated, normalization terms can be excluded for efficiency. Below is the \proglang{C} backend code for the normal distribution, where \code{distr_dim} stores the number of network statistics within the network property.

\begin{Code}
// --- Normal (distribution code == 1) ---
else if (distr_code == 1) {
    for (int i = 0; i < distr_dim; i++) {
        *g_pdf += -0.5 * pow((g_stats[i] - p1[i]), 2.0) / p2[i];
        *gp_pdf += -0.5 * pow((gp_stats[i] - p1[i]), 2.0) / p2[i];
    }
}
\end{Code}

By partitioning the extension process into these two discrete steps, \pkg{CCMnet} is an extensible package. Researchers can implement custom distributions by providing a single \proglang{R} configuration and a corresponding \proglang{C} calculation, leveraging the existing MCMC architecture for all other computational tasks.

\section{CCMnet Syntax and Diagnostics} \label{sec:syntax}

This section outlines the practical workflow for using \pkg{CCMnet}. We begin with installation requirements and the computational environment necessary to run the package. We then detail the syntax for model specification of the MCMC sampling procedure (Section~\ref{subsec:specification}). Next, we demonstrate the package's diagnostic and verification capabilities. Finally, we illustrate how the statistics of the networks generated by \pkg{CCMnet} are validated against target distributions across various network properties. Section~\ref{sec:illustrations} provides an illustrative analysis using \pkg{CCMnet} based on networks that represent friendship networks in schools.

\subsection{Installation and Requirements}

\pkg{CCMnet} is available on the Comprehensive \proglang{R} Archive Network (CRAN); version 1.1.2 or higher should be used. Because the core sampling engine is implemented in \proglang{C}, users installing from source must have a \proglang{C} compiler installed on their system (e.g., \pkg{Rtools}). The package can be installed and loaded using the following commands:

\begin{Code}
install.packages('CCMnet')
library('CCMnet')
\end{Code}

\subsection{CCM Specification and Syntax} \label{subsec:specification}

The primary function for sampling networks from a CCM in \pkg{CCMnet} is \code{sample_ccm()}. The function syntax is:

\begin{Code}
sample_ccm(network_stats, prob_distr, prob_distr_params, 
        population, cov_pattern, sample_size, burnin, interval,
        initial_g, use_initial_g, stats_only, verbose)
\end{Code}

The arguments define the CCM and the sampling parameters as follows:

\begin{itemize}
    \item \code{network_stats}: A vector of character strings specifying the network properties that define the congruence classes (e.g., \code{"edges"}, \code{"degreedist"}, \code{"degmixing"}).
    \item \code{prob_distr}: A vector of character strings defining the probability distributions placed on each property in \code{network_stats} (e.g., \code{"poisson"}, \code{"beta"}, \code{"normal"}, \code{"mvn"}).
    \item \code{prob_distr_params}: A list containing the parameters for each specified distribution.
    \item \code{population}: An integer specifying the number of nodes in the network.
    \item \code{cov_pattern}: A vector of node-level covariates, required when \code{network_stats} includes covariate-based properties.
    \item \code{sample_size}: An integer specifying the total number of networks or network statistics to store.
    \item \code{burnin}: An integer specifying the number of initial MCMC iterations to discard to ensure the sampler has reached the target distribution before a network is generated.
    \item \code{interval}: An integer specifying the MCMC thinning interval between stored network data or statistics.
    \item \code{initial_g}: An optional \pkg{igraph} graph object providing the starting state for the MCMC algorithm.
    \item \code{use_initial_g}: A logical value; if \code{TRUE}, the sampler begins at \code{initial_g}. If \code{FALSE}, the sampler generates its own starting graph.
    \item \code{stats_only}: A logical value; if \code{TRUE}, only the MCMC chain of network statistics is returned and final network. If \code{FALSE}, the function returns a list of networks, where each is an individual \code{igraph} object (from the \pkg{igraph} package).
    \item \code{verbose}: A integer specifying the level of output logs (0 = silent, 1 = basic, 2 = detailed).
\end{itemize}

The function \code{sample_ccm()} returns an S3 object of class \code{ccm_sample}. This object is a list containing the generated network(s) as a list (\code{g}), where each network is represented as an \pkg{igraph} object \citep{igraph_rcran}. The object also contains the matrix of MCMC sampled statistics (\code{mcmc_stats}) and the model metadata (population size, probability distributions, and covariates). Users can apply standard \pkg{igraph} functions, such as \code{plot()} or \code{degree()}, to the resulting network(s). Furthermore, to facilitate standard \proglang{R} workflows, \pkg{CCMnet} provides several methods for interacting with \code{ccm_sample} objects:

\begin{itemize} 
    \item \code{print(x, \ldots)}: Displays a concise summary of the model configuration, including the targeted network properties and the number of MCMC samples collected. 
    \item \code{summary(object, \ldots)}: Provides descriptive statistics (minimum, maximum, mean, and quartiles) for each sampled network property in the MCMC chain. 
    \item \code{plot(x, stats, type, target_distr, \ldots)}: Generates visualizations of the empirical distributions derived from the MCMC samples. The supported plot types include histograms (\code{``hist''}), density plots (\code{``density''}), and trace plots (\code{``trace''}). 
\end{itemize}

\subsection{Diagnostics and Verification: Conceptual}

Before using a generated network ensemble for analysis, it is important to investigate that the MCMC sampler has converged and that the resulting empirical distribution matches the target distribution specification $P_{\phi}(x|\theta)$ defined in the \code{sample_ccm}. \pkg{CCMnet} provides two specialized functions for these purposes:

\begin{itemize} 
    \item \code{sample_target_distr(ccm_sample, n_sim)}: Facilitates statistical verification by drawing \code{n_sim} independent samples directly from the target distributions defined in the \code{sample_ccm} object. Unlike the MCMC algorithm, which generates actual graph structures, this function draws vectors of real values representing potential network summary statistics. Therefore, these draws serve as a reference to see if the network summary statistics from the networks generated by the MCMC algorithm align with the target distribution for the summary statistics.
    \item \code{plot(ccm_sample, target_distr = TRUE)}: The \code{plot} method generates plots for the specified network statistics. When \code{target_distr = TRUE}, the function allows the user to visually inspect the convergence, alignment, and mixing behavior of the MCMC chain in comparison to the target distribution and determine if adjustments are necessary.
\end{itemize}

Once the  target probability distribution samples are generated, they can be visualized alongside the MCMC output using the \code{plot()} method:

\begin{Code}
ccm_sample <- sample_target_distr(ccm_sample, n_sim = 1000)
plot(ccm_sample, stats, type, target_distr = TRUE).
\end{Code}

When \code{target_distr = TRUE}, the \code{plot} method overlays the target samples (or quantiles in trace plots) on the MCMC results. Close agreement between the MCMC-derived densities and the target densities provides a direct diagnostic of the alignment between the summary statistics of networks generated by the MCMC algorithm and the target probability distribution for these statistics. This verification step is a core feature of the \pkg{CCMnet} workflow, ensuring that the target distribution on the network properties specified by the user align with the empirical distribution from the generated networks.

\subsection{Diagnostics and Verification: Simple Example}

To demonstrate the distinction between a target distribution and the realized distribution over the space of graphical networks, we provide a small case study for illustration where the number of nodes is $n=4$. This example with a small number of nodes allows for the exact enumeration of all graphical network property values, demonstrating the importance of comparing the target distribution with the empirical distribution.

\subsubsection{Support Mismatch and Graphical Constraints}

In this example, we define the network statistic of interest as the degree distribution. For a network of size $n=4$, there are $35$ possible partitions of the number of nodes into four values (ranging from 0 to 4). However, the structural constraints dictate that only $11$ of these partitions are graphical degree distributions, i.e., values that can be realized by at least one simple graph.

\begin{figure}
\centering
\includegraphics[width=0.9\textwidth]{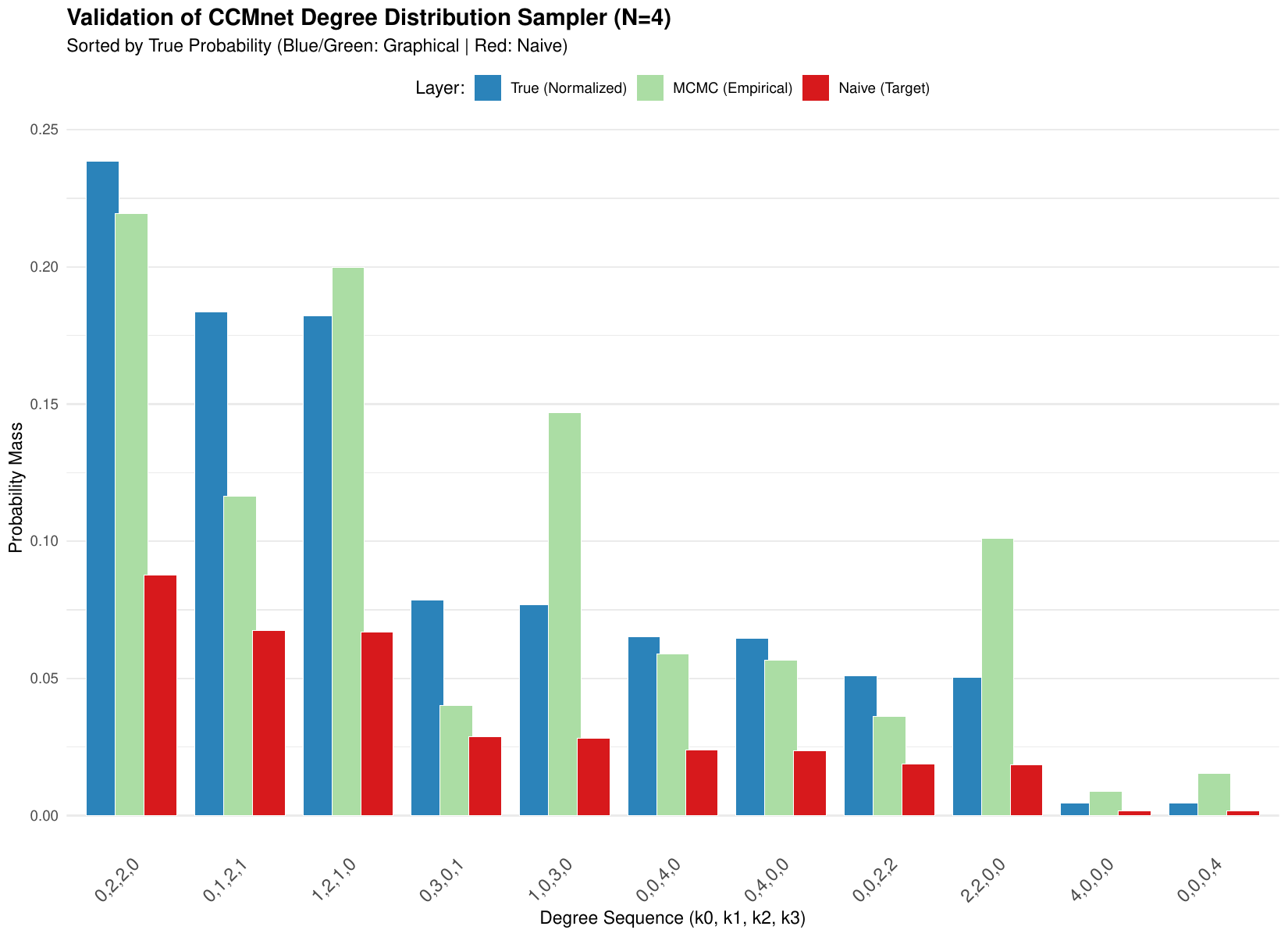}
\caption{Comparison of probability distributions for $N=4$. The target distribution ($Q_{\phi}(x \vert \theta)$) is shown by the red bars. The normalized probability for the 11 graphical degree sequences ($P_{\phi}(x \vert \theta)$) is shown by the blue bars. The green bars show the MCMC samples from \pkg{CCMnet}. }
\label{fig:CCM_ex_4_node.}
\end{figure}

We specify a Dirichlet-Multinomial distribution as our target density, with the target distribution denoted as $Q_{\phi}(x \vert \theta)$. The red bars in Figure~\ref{fig:CCM_ex_4_node.} represent the values of $Q_{\phi}(x \vert \theta)$ for the 11 graphical sequences. Crucially, because the full support of $Q_{\phi}(x \vert \theta)$ includes all 35 partitions---24 of which are non-graphical---the red bars do not sum to $1$. 

\subsubsection{Renormalization to Graphical Support}

When the set of graphical degree distributions is known, as is the case for $N=4$, we can re-normalize the density to the graphical support. We denote the resulting restricted probability on graphical values as $P_{\phi}(x \vert \theta)$, where $P_{\phi}(x \vert \theta) \propto Q_{\phi}(x \vert \theta)$ for all graphical $x$, and $P_{\phi}(x \vert \theta) = 0$ otherwise.

The blue bars represent this probability $P_{\phi}(x \vert \theta)$. The reason why the original target distribution $Q_{\phi}(x \vert \theta)$ differs from the restricted probability $P_{\phi}(x \vert \theta)$ is due to the mass lost to non-graphical values. 

\subsubsection{Performance and Asymptotics}

Ideally, the \pkg{CCMnet} MCMC should be sampling from $P_{\phi}(x \vert \theta)$. The MCMC empirical distribution (green bars) represent the frequency of states actually visited by the sampler. As seen in Figure~\ref{fig:CCM_ex_4_node.}, the blue and green bars are clearly more similar than the blue and red bars. This confirms that the sampler is correctly targeting the restricted probability $P_{\phi}(x \vert \theta)$ rather than the unnormalized $Q_{\phi}(x \vert \theta)$.

The discrepancies remaining between $P_{\phi}(x \vert \theta)$ (blue) and the MCMC sample (green) bars are likely due to the fact that the congruence class ratio estimates are based on large-graph asymptotic approximations. These asymptotics are not yet realized at $n=4$. Nevertheless, the sampler demonstrates high fidelity in capturing the relative weights of the graphical distribution, even under these restricted conditions. Section~\ref{sec:illustrations} demonstrates align between the target distribution and MCMC samples for larger values of $n$. 

\section{Illustrative Examples} \label{sec:illustrations}

We demonstrate the utility of \pkg{CCMnet} through a series of examples using topology-based specifications, ranging from simple edge counts to complex joint degree distributions with clustering. Each example follows a consistent workflow: model specification, MCMC sampling, verification, and generate network ensemble. Verification is performed by comparing the empirical distribution of network properties obtained from the MCMC to the corresponding target distribution. These examples serve both as a guide for users to construct their own models and to validate the software implementation of the MCMC theory that underpins  CCMs across different congruence class definitions. Following these example demonstrations, Section~\ref{sec:illustrations} provides an applied analysis using school friendship networks to illustrate the package's utility in a real-world research context.

\subsection{Edge Count}

We first consider a CCM in which a soft constraint is placed on the total number of edges. In particular, suppose the number of edges follows a Poisson random variable with mean $\lambda = 350$. Under this model, the probability of sampling a network with $k$ edges is:

\begin{equation}
    P(k \mid \lambda) = \frac{\lambda^k e^{-\lambda}}{k!}.
\end{equation}

The CCM assigns this probability to the congruence class consisting of all networks with $k$ edges and assigns equal probability to each network within that class. Using the following specification of \code{sample_ccm}, we sample a network from this model:

\begin{Code}
ccm_sample <- sample_ccm(
  network_stats = c("edges"),
  prob_distr = c("poisson"),
  prob_distr_params = list(list(350)), 
  population = 50L
)
\end{Code}

Note that the underlying network generated by this process is \code{ccm_sample$g[[1]]} and it is stored as an \code{igraph} object within \code{ccm_sample}, allowing for immediate visualization or calculation of additional metrics. Use the \code{summary} and \code{print} commands to get information about the \code{ccm_sample} object. 

\begin{CodeChunk}
\begin{CodeInput}
R> summary(ccm_sample)
\end{CodeInput}
\begin{CodeOutput}
Summary of ccm_sample object
-------------------------

Statistic:  edges 
   Min. 1st Qu.  Median    Mean 3rd Qu.    Max. 
  292.0   336.0   349.0   348.3   361.0   411.0 
\end{CodeOutput}
\begin{CodeInput}
R> print(ccm_sample)
\end{CodeInput}
\begin{CodeOutput}
Object of class 'ccm_sample'
-------------------------
Statistics:        edges 
Distribution(s):   poisson 
Population:        50 
MCMC samples:      1000 rows x 1 cols
\end{CodeOutput}
\end{CodeChunk}

To verify the sampler, we compare the empirical distribution of edge counts from the MCMC output to samples drawn directly from the Poisson distribution using \code{sample_target_distr()}. Successful verification is indicated by close agreement between the two distributions, visualized using the \code{plot()} method:

\begin{Code} 
ccm_sample <- sample_target_distr(ccm_sample, n_sim = 1000)
plot(ccm_sample, stats = "edges", type = "hist", target_distr = TRUE)
\end{Code}

\begin{figure}[t!]
    \centering
    \begin{subfigure}[b]{0.32\textwidth}
        \centering
        \includegraphics[width=\linewidth]{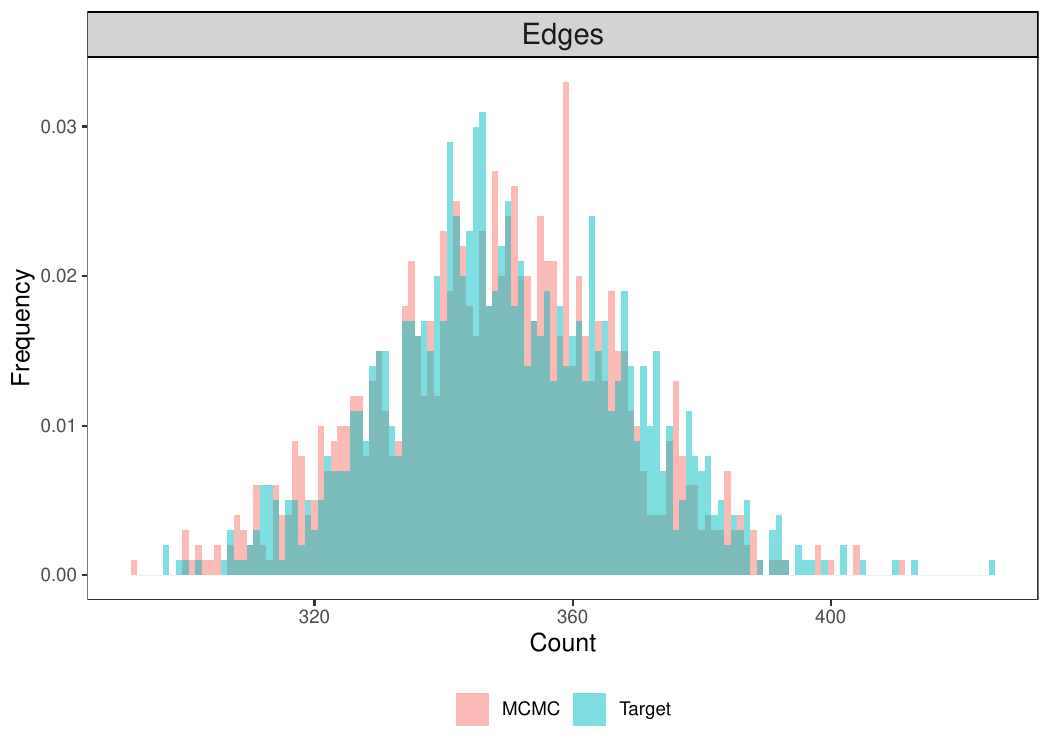}
        \caption{Poisson}
        \label{fig:edges_a}
    \end{subfigure}
    \hfill
    \begin{subfigure}[b]{0.32\textwidth}
        \centering
        \includegraphics[width=\linewidth]{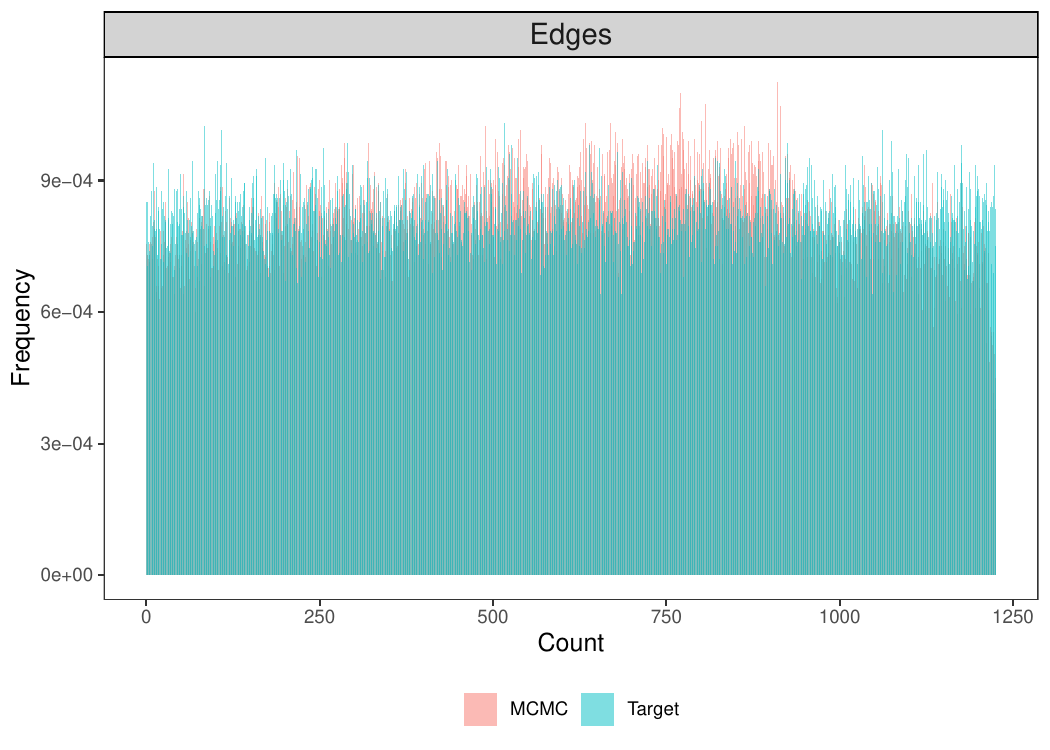}
        \caption{Uniform}
        \label{fig:edges_b}
    \end{subfigure}
    \hfill
    \begin{subfigure}[b]{0.32\textwidth}
        \centering
        \includegraphics[width=\linewidth]{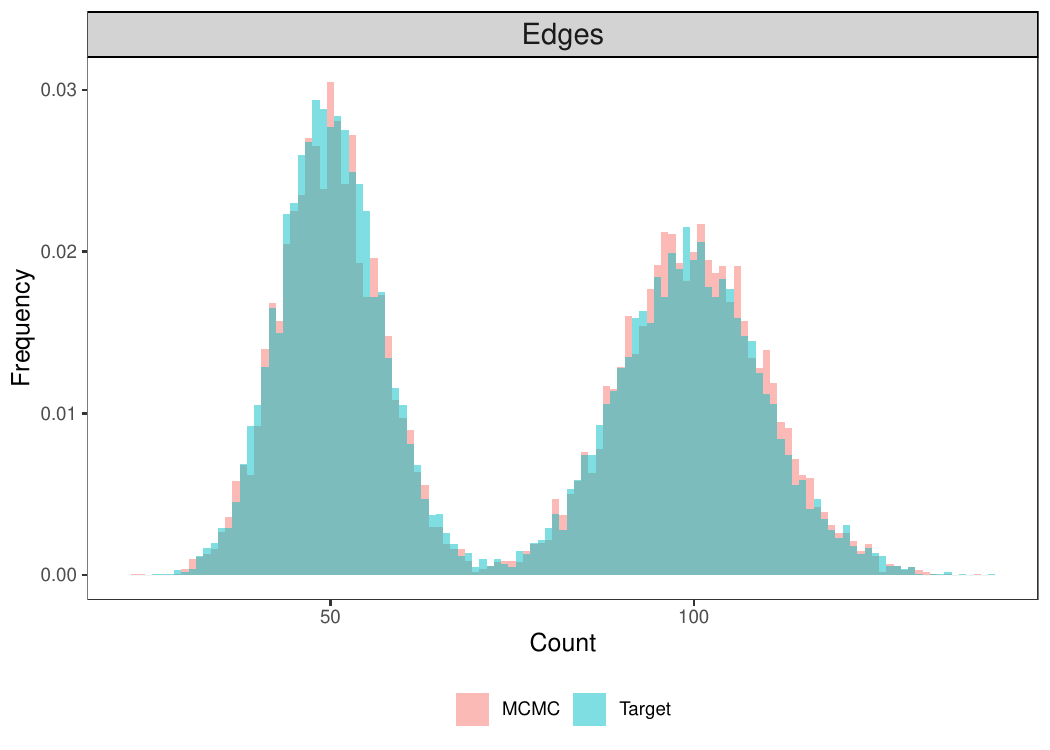}
        \caption{Nonparametric}
        \label{fig:edges_c}
    \end{subfigure}
    \caption{Distribution of network edge counts for three CCM specifications: (a) Poisson distribution with $\lambda = 350$; (b) Uniform distribution over the support $[0, {N \choose 2}]$; and (c) Bimodal nonparametric distribution proportional to the sum of Poisson$(\lambda = 50)$ and Poisson$(\lambda = 100)$. Red bars represent the empirical density of MCMC samples, while blue bars represent independent draws from the target distributions.}
    \label{fig:CCM_ex_edges}
\end{figure}

In Figure~\ref{fig:CCM_ex_edges} (Panel A), the empirical distribution obtained from the MCMC sampler (red) closely matches the target distribution (blue), confirming correct sampler behavior.

\pkg{CCMnet} also supports simpler or more complex distributions for number of edges. For example, a uniform distribution over edge counts is achieved by setting \code{prob_distr = c("uniform")} and \code{prob_distr_params = NULL}. To ensure convergence to the stationary distribution and enable validation, we set a sufficient \code{burnin} and \code{sample_size}. The verification result is shown in Panel B of Figure~\ref{fig:CCM_ex_edges}.

\begin{Code}
ccm_sample <- sample_ccm(
  network_stats = c("edges"),
  prob_distr = c("uniform"), 
  prob_distr_params = list(NULL),
  population = 50L,
  sample_size = 200000L,
  burnin = 100000L
)

ccm_sample<- sample_target_distr(ccm_sample, n_sim = 200000)

plot(ccm_sample, stats = "edges", type = "hist", target_distr = TRUE)
\end{Code}

More generally, arbitrary edge count distributions can be specified using the nonparametric (\code{"NP"}) option. In this case, the user provides a vector $\alpha$ of length ${N \choose 2} + 1$, where $\alpha_k$ denotes the probability of sampling a network with $k$ edges. We illustrate this flexibility using a bimodal specification where $\alpha$ is proportional to the sum of two Poisson densities, Poisson$(\lambda=50)$ and Poisson$(\lambda=100)$. The verification results are shown in Panel C of Figure~\ref{fig:CCM_ex_edges}.

\begin{Code}
n_max <- choose(50, 2)
alpha <- dpois(0:n_max, lambda = 50) + dpois(0:n_max, lambda = 100)
prob_distr_params <- alpha / sum(alpha)

ccm_sample <- sample_ccm(
  network_stats = c("edges"),
  prob_distr = c("np"),
  prob_distr_params = list(list(prob_distr_params)),
  population = 50L,
  sample_size = 10000L,
  burnin = 100000L
)

ccm_sample<- sample_target_distr(ccm_sample, n_sim = 10000)

plot(ccm_sample, stats = "edges", type = "hist", target_distr = TRUE)
\end{Code}

\subsection{Degree Distribution}

Next, we illustrate the \pkg{CCMnet} sampler for models specified by the degree distribution. Here, the network property of interest is the frequency of each degree within the network. We model this distribution using a Dirichlet-Multinomial distribution, which allows for flexible modeling of degree frequencies while accounting for overdispersion.

\begin{Code}
ccm_sample<- sample_ccm(network_stats = c('degreedist'),
                        prob_distr = c('dirmult'),
                        prob_distr_params = list(list(c(2,21,15,12))), 
                        population = 100L, 
                        sample_size = 10000L,
                        burnin = 100000L) 

ccm_sample <- sample_target_distr(ccm_sample, n_sim = 10000)

plot(ccm_sample, 
     stats = paste0("deg", 0:3), 
     type = "hist", 
     target_distr = TRUE) 
\end{Code}

\begin{figure}
\centering
\includegraphics[width=\textwidth]{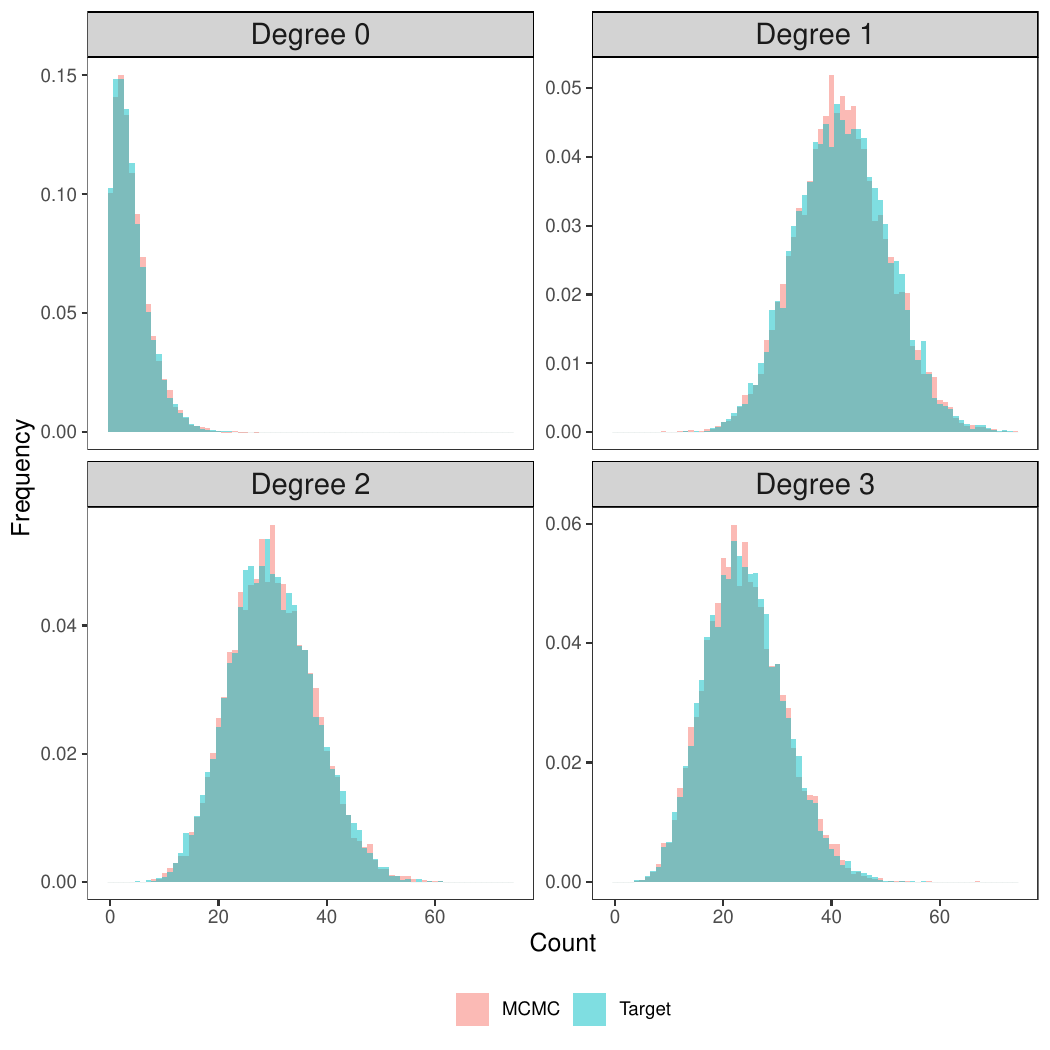}
\caption{Node degree distributions for a Dirichlet-Multinomial CCM specification. Panels represent the frequency of nodes with degrees 0 through 3. Red bars represent the empirical histogram of MCMC samples, while blue bars represent independent draws from the target Dirichlet-Multinomial distribution}
\label{fig:CCM_ex_degree}
\end{figure}

Figure~\ref{fig:CCM_ex_degree} displays the resulting density plots for the number of nodes with degrees $0$ through $3$. Red bars show an empirical histogram from the MCMC samples, while blue bars show a histogram obtained from the target distribution.

\subsection{Degree Mixing and Clustering}

Finally, we illustrate the \pkg{CCMnet} sampler for models involving higher-order dependencies, specifically degree mixing and the total number of triangles. In this example, we place a multivariate normal specification on the degree mixing matrix and a univariate normal, $N(\mu=10, \sigma^2=3)$,  specification on the triangle count. Because the degree mixing matrix effectively determines the number of connected triples, the simultaneous specification on triangle counts implicitly controls the network's clustering.

\begin{Code}
mean_vec = c(23, 66, 44, 20, 120, 80)
var_mat = matrix(data = c(22, -3, -2, -5, -6, -4,
                          -3, 58, -7, -14, -18, -12,
                          -2, -7, 41, -9, -12, -8,
                          -5, -14, -9, 75, -25, -17,
                          -6, -18, -12, -25, 89, -22,
                          -4, -12, -8, -17, -22, 68), ncol = 6)
prob_distr_params = list(list(mean_vec, var_mat),
                         list(10,3))

ccm_sample <- sample_ccm(network_stats = c('degmixing', 'triangles'),
                         prob_distr = c('mvn', 'normal'),
                         prob_distr_params = prob_distr_params,
                         population = 500,
                         sample_size = 100000,
                         burnin = 500000) 

ccm_sample <- sample_target_distr(ccm_sample, n_sim = 100000)

plot(ccm_sample, 
     stats = c("DM11", "DM12", "DM13", "DM22", "DM23", "DM33", "triangles"), 
     type = "hist", 
     target_distr = TRUE)

plot(ccm_sample, 
     stats = c("triangles"), 
     type = "trace", 
     target_distr = TRUE)
\end{Code}

\begin{figure}[t!] 
\centering 
\includegraphics[width=\textwidth]{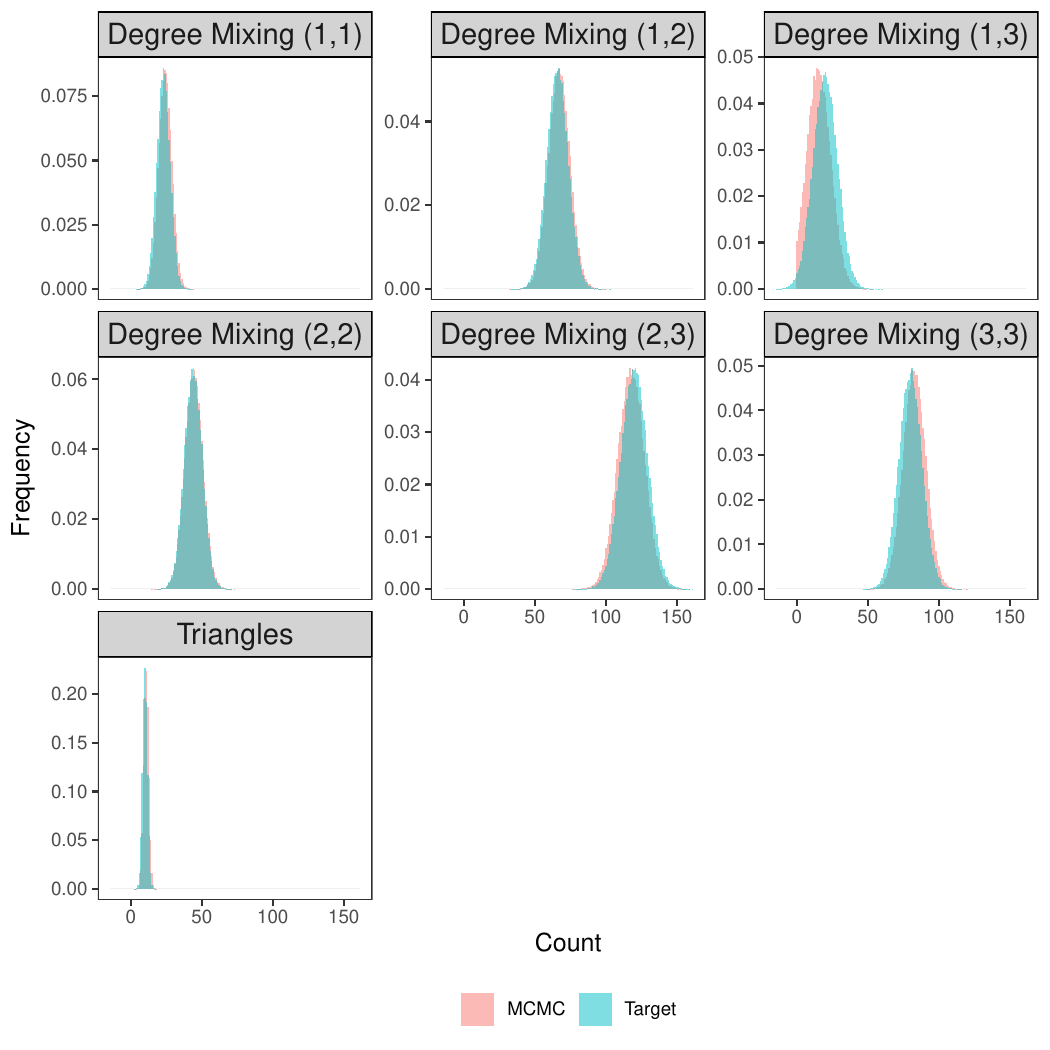} 
\caption{Distribution of degree mixing entries and triangle counts. Red bars represent empirical MCMC frequencies; blue bars represent target distributions. The degree mixing entries are indexed by the degrees of the connecting nodes.} \label{fig:CCM_ex_degmixing_triangle} 
\end{figure}

\begin{figure}[t!] 
\centering 
\includegraphics[width=\textwidth]{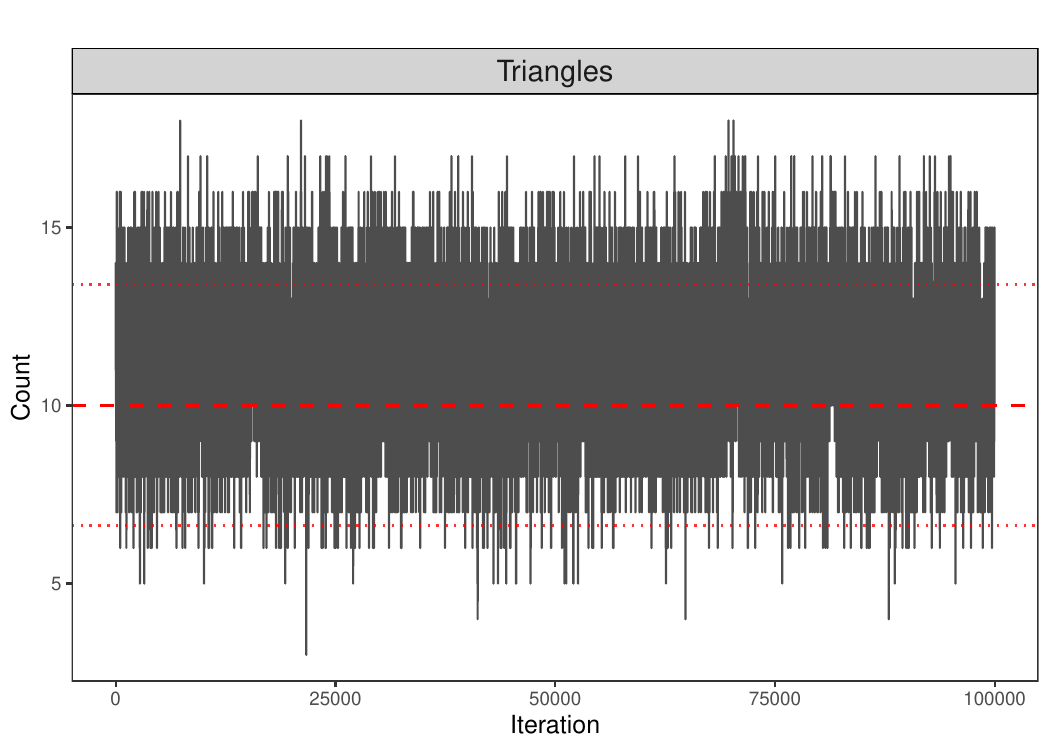} 
\caption{MCMC trace plot for the triangle count statistic, illustrating the mixing behavior and stationarity of the sampler under joint topological constraints. The horizontal dashed line represents the target mean, and the dotted lines represent the 2.5\% and 97.5\% quantiles of the target distribution.} 
\label{fig:CCM_ex_degmixing_triangle_traceplot} 
\end{figure}

Figure~\ref{fig:CCM_ex_degmixing_triangle} displays the resulting histograms for the unique entries of the degree mixing matrix and the triangle count. While the histograms confirm that the sampler reaches the correct stationary distribution, the trace plot in Figure~\ref{fig:CCM_ex_degmixing_triangle_traceplot} provides further diagnostic evidence of mixing and stationarity. By setting \code{target_distr = TRUE}, the trace plot includes a horizontal dashed line representing the target mean and dotted lines for the 2.5\% and 97.5\% quantiles. The close alignment of the MCMC chain with these reference benchmarks demonstrates that the sampler effectively explores the high-dimensional space defined by the joint topological targets.

\section{Illustrative Application: School Friendship Networks} \label{sec:application}

In this section, we provide a comprehensive workflow for investigating the spread of a communicable disease within a school setting. The utility of \pkg{CCMnet} lies in its ability to generate network ensembles that reflect both the structure and the inherent uncertainty of social contacts, which results in more robust estimates for, say, subsequent epidemic forecasting. This analytical workflow proceeds through five stages, where the middle three represent the core components associated with CCMs:

\begin{enumerate}

    \item \textbf{Network Data identification:} Network data can be extracted from existing historical repositories or obtained via a new empirical study. Regardless, they are typically obtained through sampling. The choice of sampling design is contingent on the population of interest and dictates the downstream estimation process, including which network properties can be estimated and the appropriate distributional assumptions for the observed data. 
    
    \item \textbf{Network property specification:} Translating observed data and theoretical assumptions into target properties (e.g., density, degree distribution) and their associated probability distributions.

    \item \textbf{Validation:} Ensuring that the MCMC sampler has successfully converged to the expected target distribution.

    \item \textbf{Network generation:} Producing the final ensemble of networks for follow-on simulations.

    \item \textbf{Epidemic simulation:} Leveraging the generated network ensembles to investigate network-based scientific and epidemiological questions.
    
\end{enumerate}

\subsection{Network Data Identification}

For our illustration, data are collected by sampling individuals (nodes) from a single school friendship network. Common single-network sampling designs include egocentric sampling \citep{marsden1990network} and respondent-driven sampling \citep{heckathorn1997respondent}, both of which have been widely applied to investigate social and sexual partnerships \citep{laumann1999racial, krivitsky2017inference}. For this workflow, we consider the Dixon High School network, which is one of four school friendship networks (Mesa, Desert, Dixon, and Magnolia) simulated to match the structural characteristics of schools in the National Longitudinal Study of Adolescent Health and included in the \pkg{statnet} suite of packages \citep{Handcock_2018_statnet, HunterHandcock2008}.

We assume that the complete Dixon network represents the full population and friendships. The network environment can be initialized and verified using the following commands:

\begin{Code} 
utils::data("faux.dixon.high", package = "ergm", envir = environment())
dixon_net = intergraph::asIgraph(faux.dixon.high)
\end{Code}

\begin{CodeChunk}
\begin{CodeInput}
R> igraph::vcount(dixon_net)
\end{CodeInput}
\begin{CodeOutput}
248
\end{CodeOutput}
\begin{CodeInput}
R> igraph::gsize(dixon_net)
\end{CodeInput}
\begin{CodeOutput}
1197
\end{CodeOutput}
\end{CodeChunk}

\noindent As shown above, the complete network contains $248$ individuals (nodes) and $1,197$ friendships (edges). To simulate partial observation of the school environment, subsets of nodes were randomly sampled across a range of sample sizes $n \in \{25, 50, \dots, 225\}$. For each sampled node, we assume for this illustration that only the number of friendships associated with that individual (degree) is recorded.

\subsection{Network Property Specification}

For this illustration, we are interested in simulating networks with the same network density as the Dixon network. To do so, we estimate a posterior distribution for network density based on the partially observed network data. Note that while it is possible to model other structural network features, such as the degree distribution, for simplicity we restrict our investigation to network density.

For our analysis, we assume a homogeneous tie-formation propensity; that is, each friendship is formed with the same probability regardless of the presence or absence of other network edges. Therefore, we model the observed edges using a Bernoulli likelihood paired with a conjugate Beta prior.

For each sample size, the ratio of observed edges to potential edges provides the likelihood for the edge propensity. Using a non-informative Beta$(1,1)$ prior, we estimate the posterior distribution of the network density, applying a finite population correction to the posterior variance. We utilize the \code{postmix} function from the \pkg{RBesT} package to compute the posterior parameters \citep{RBesT_rcran}. The code below estimates the posterior distribution for network density when $100$ nodes are sampled from Dixon High School.

\begin{Code}
dixon_deg = igraph::degree(dixon_net)
prior.unif <- RBesT::mixbeta(c(1, 1, 1))
N = igraph::vcount(dixon_net)

num_sample_nodes = 100

dixon_deg_sample = sample(x = dixon_deg, size = num_sample_nodes, replace = FALSE)

r=sum(dixon_deg_sample)
n=sum(num_sample_nodes*(N-1))
posterior.sum_beta <- RBesT::postmix(prior.unif, 
                                     n=n, 
                                     r=r)

alpha_post <- posterior.sum_beta[2]
beta_post  <- posterior.sum_beta[3]

# Infinite-population variance
var_inf <- (alpha_post * beta_post) /
  ((alpha_post + beta_post)^2 * (alpha_post + beta_post + 1))

# Finite population correction
fpc <- (N - num_sample_nodes) / (N - 1)
var_fpc <- var_inf * fpc

# Moment-matched Beta parameters
mu <- alpha_post / (alpha_post + beta_post)
S <- mu * (1 - mu) / var_fpc - 1
posterior.sum_beta[2] <- mu * S
posterior.sum_beta[3] <- (1 - mu) * S
\end{Code}

\begin{figure}[t!]
\centering
\includegraphics[width=\textwidth]{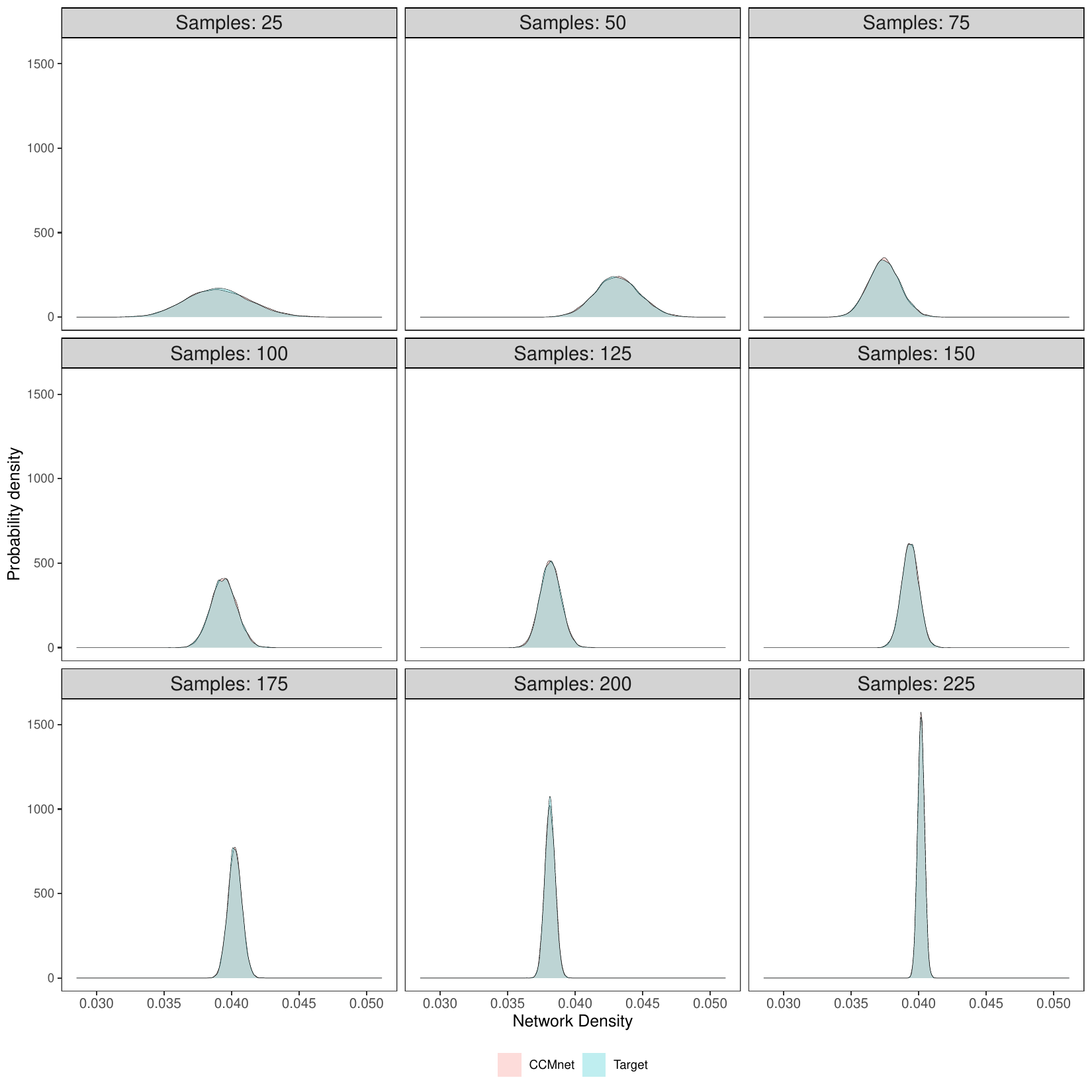}
\caption{Network density distributions for within-network sampling across varying number of individuals sampled ($n \in \{25, 50, \dots, 225\}$) from the school. Within each plot, network statistic distributions for the CCM (red) and the target distribution (blue) are presented.}
\label{fig:CCM_ex_density_within_school_validate}
\end{figure}

Figure~\ref{fig:CCM_ex_density_within_school_validate} (blue curves) represents the samples from the estimated posterior distribution for sample sizes $n \in \{25, 50, \dots, 225\}$.

\subsection{Validation}

Misalignment between the summary statistics of networks generated by \pkg{CCMnet} and the target posterior distribution can result from numerical instability, insufficient number of MCMC draws, or differences between the target distribution and graphical network summary statistics. The purpose of validation is to diagnose and quantify the extent of this potential misalignment prior to generating the network ensemble for analysis.

Validation consists of comparing the empirical distribution of network density from networks generated by the MCMC to samples drawn directly from the target posterior distribution using \code{sample_target_distr()}. The code below generates the empirical distribution from \pkg{CCMnet} using the posterior distribution (estimated in Step 2) as the target distribution:

\begin{Code}
n_samples <- 10000L

ccm_sample <- sample_ccm(
  network_stats = c("density"),
  prob_distr = c("beta"),
  prob_distr_params = list(list(posterior.sum_beta[2], posterior.sum_beta[3])), 
  population = N,
  sample_size = n_samples,
  burnin = 100000L
)

# Extract reference samples directly from the target distribution
ccm_sample <- sample_target_distr(ccm_sample, n_sim = n_samples)
\end{Code}

For this illustration, the empirical distribution obtained from the MCMC sampler (red curves) closely matches the target distribution (blue curves), confirming correct MCMC sampler behavior across all evaluated sample sizes (Figure~\ref{fig:CCM_ex_density_within_school_validate}). It is worth noting that while the samples drawn directly from the posterior distribution represent real-valued network summary statistics rather than discrete graphs, the sampler successfully constrains the discrete network configurations to honor these continuous distributions.

\subsection{Generating Network Ensembles} \label{sec:ensemble}

Once the MCMC chain has been verified for stationarity and the simulated summary statistics align with the target distribution, an ensemble of networks can be generated for subsequent analysis. For the school friendship networks described above, we first perform a warm-up run to ensure convergence (see code in Validation step), then utilize the \code{use_initial_g},  \code{initial_g}, and \code{stats_only} parameters to efficiently produce the final ensemble.

\begin{Code}
school_ensemble <- sample_ccm(
  network_stats = c("density"),
  prob_distr = c("beta"),
  prob_distr_params = list(list(posterior.sum_beta[2], posterior.sum_beta[3])), 
  population = N,
  sample_size = 1000L,
  burnin = 1L,
  interval = 1000L,
  initial_g = ccm_sample$g[[1]], 
  use_initial_g = TRUE, 
  stats_only = FALSE)
\end{Code}

In the above call, setting \code{use_initial_g = TRUE} and passing the final network state from the initial run (\code{initial_g = ccm_sample$g[[1]]}) ensures the sampler begins directly on the target manifold. By setting \code{burnin = 1}, we avoid unnecessary computation, and setting \code{stats_only = FALSE} ensures that the networks are retained. The resulting network ensemble is stored in \code{school_ensemble$g} as an R \code{list} of \code{igraph} objects, as demonstrated below:

\begin{CodeChunk}
\begin{CodeInput}
R> class(school_ensemble$g)
\end{CodeInput}
\begin{CodeOutput}
[1] "list"
\end{CodeOutput}
\begin{CodeInput}
R> length(school_ensemble$g)
\end{CodeInput}
\begin{CodeOutput}
[1] 1000
\end{CodeOutput}
\begin{CodeInput}
R> class(school_ensemble$g[[1]])
\end{CodeInput}
\begin{CodeOutput}
[1] "igraph"
\end{CodeOutput}
\end{CodeChunk}

These 1,000 networks can now be used to investigate disease dynamics in schools.

\subsection{Simulate Epidemic}

In this section, we demonstrate how to use the generated networks to simulate disease dynamics. For our school-based influenza scenario, we pass the network list through the stochastic Susceptible-Infectious-Recovered (SIR) simulation framework provided by the \pkg{igraph} package using the function \code{sir}. This allows us to capture two distinct levels of variation: 1) the stochastic randomness of disease transmission on any single contact structure, and 2) the structural variance across different network configurations within the ensemble.

To benchmark our findings, we compare the CCM ensemble to the $G(n,m)$ model matching the expected mean edge count of the posterior distribution. The complete simulation is executed as follows:

\begin{Code}
# 1. Generate baseline G(n,m) random networks matching expected density
p_density <- posterior.sum_beta[2] / (posterior.sum_beta[2] + posterior.sum_beta[3])
N <- igraph::vcount(dixon_net)
m <- round(choose(N, 2) * p_density)

school_ensemble_er <- list()
for (i in 1:length(school_ensemble$g)) {
  school_ensemble_er[[i]] <- igraph::sample_gnm(n = N, m = m, directed = FALSE)
}

# Combine the CCM and baseline ensembles
school_ensemble_comb <- c(school_ensemble$g, school_ensemble_er)

# 2. Define disease parameters and run stochastic SIR models
transmission_rate <- 0.075  
recovery_rate     <- 0.300  
sims_per_network  <- 1L     

sir_ensemble_results_comb <- lapply(school_ensemble_comb, function(graph) {
  igraph::sir(graph, beta = transmission_rate, gamma = recovery_rate, 
              no.sim = sims_per_network)
})

# 3. Helper function to parse individual sir objects and extract metrics
extract_epidemic_metrics <- function(sir_obj, num_nodes) {
  metrics <- lapply(sir_obj, function(run) {
    max_infected    <- max(run$NI)
    total_recovered <- tail(run$NR, 1) 
    attack_rate     <- total_recovered / num_nodes
    return(data.frame(PeakInfected = max_infected, AttackRate = attack_rate))
  })
  return(do.call(rbind, metrics))
}

# Collapse simulations across all network architectures into a data frame
num_nodes <- igraph::vcount(school_ensemble$g[[1]])
combined_results <- do.call(rbind, lapply(sir_ensemble_results_comb, 
                                          extract_epidemic_metrics, 
                                          num_nodes = num_nodes))

combined_results$sim <- c(rep("CCM", length(school_ensemble$g)), 
                          rep("ER", length(school_ensemble_er)))
\end{Code}

\begin{figure}[t!]
\centering
\includegraphics[width=\textwidth]{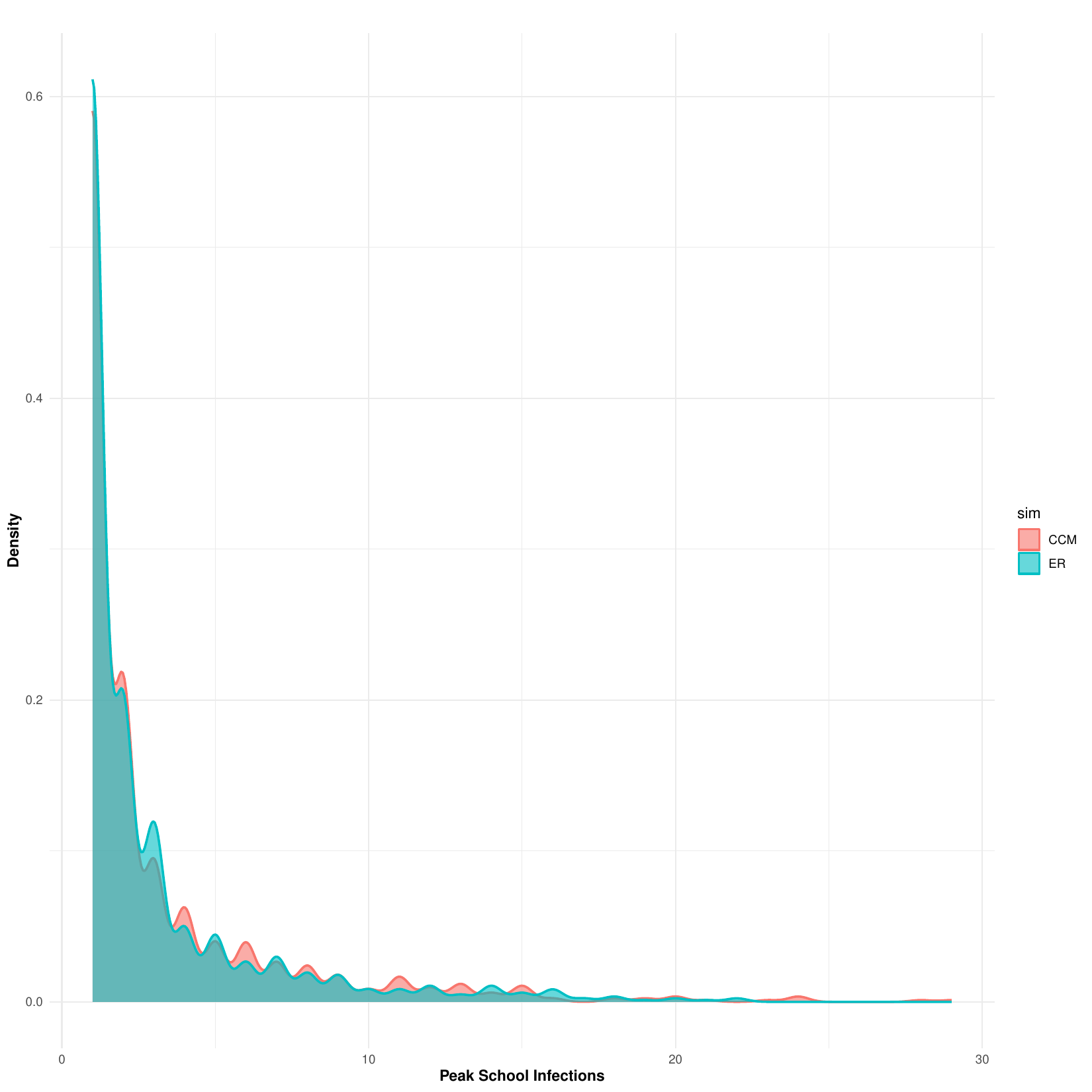}
\caption{Distribution of simulated epidemic outcomes across the network ensemble based on CCMs (red curve) and the $G(n,m)$ model (blue curve). The x-axis represents the peak number of students infected at a given time of the simulation, and the y-axis represents the density.}
\label{fig:CCM_ex_density_within_school_sim}
\end{figure}

By extracting the number of infections over time, researchers can compute the mean, median, and 95\% uncertainty intervals for key public health metrics, such as the peak infection size and the total infections. Figure~\ref{fig:CCM_ex_density_within_school_sim} shows the distribution (red curve) of the peak number of students infected at a given time of the simulation across the network ensemble. 

In addition, the figure shows the distribution obtained when the networks are generated based on the benchmark $G(n,m)$ model (blue curve). A comparison between these modeling approaches indicates that incorporating uncertainty using CCMs alters projections for the peak number of infections. In this case, the CCM approach yields an approximately 7\% higher mean peak number of infections, demonstrating the potential importance of integrating uncertainty into the modeling.

\subsection{Comparison}

In this section, we demonstrate how \pkg{CCMnet} captures the varying degrees of uncertainty inherent in real-world network data. Specifically, we compare the resulting posterior predictive distributions to networks generated under two alternative benchmark models: the $G(n,m)$ model and an ERGM with a single term for edges. The $G(n,m)$ model imposes a fixed constraint on the number of edges, while the ERGM imposes a soft constraint with a fixed level of variability. 

For our comparison, the log-odds edge coefficient ($\theta$) for the ERGM is set to correspond directly to the calculated mean density of the estimated posterior, and $m$ for the $G(n,m)$ model is set to the exact expected mean edge count. The ERGM framework is initialized and simulated using the following framework:

\begin{Code}
# Initialize an empty network object matching the population size
net <- network::network(N, directed = FALSE)

# Compute the expected edge propensity and convert to log-odds (theta)
p <- posterior.sum_beta[2] / (posterior.sum_beta[2] + posterior.sum_beta[3])
theta <- log(p / (1 - p))

# Simulate edge statistics from the baseline ERGM structure
ERGM <- simulate(
  net ~ edges,
  coef = theta,
  nsim = n_samples,
  output = "stats"
) / choose(N, 2)
\end{Code}

\begin{figure}[t!]
\centering
\includegraphics[width=\textwidth]{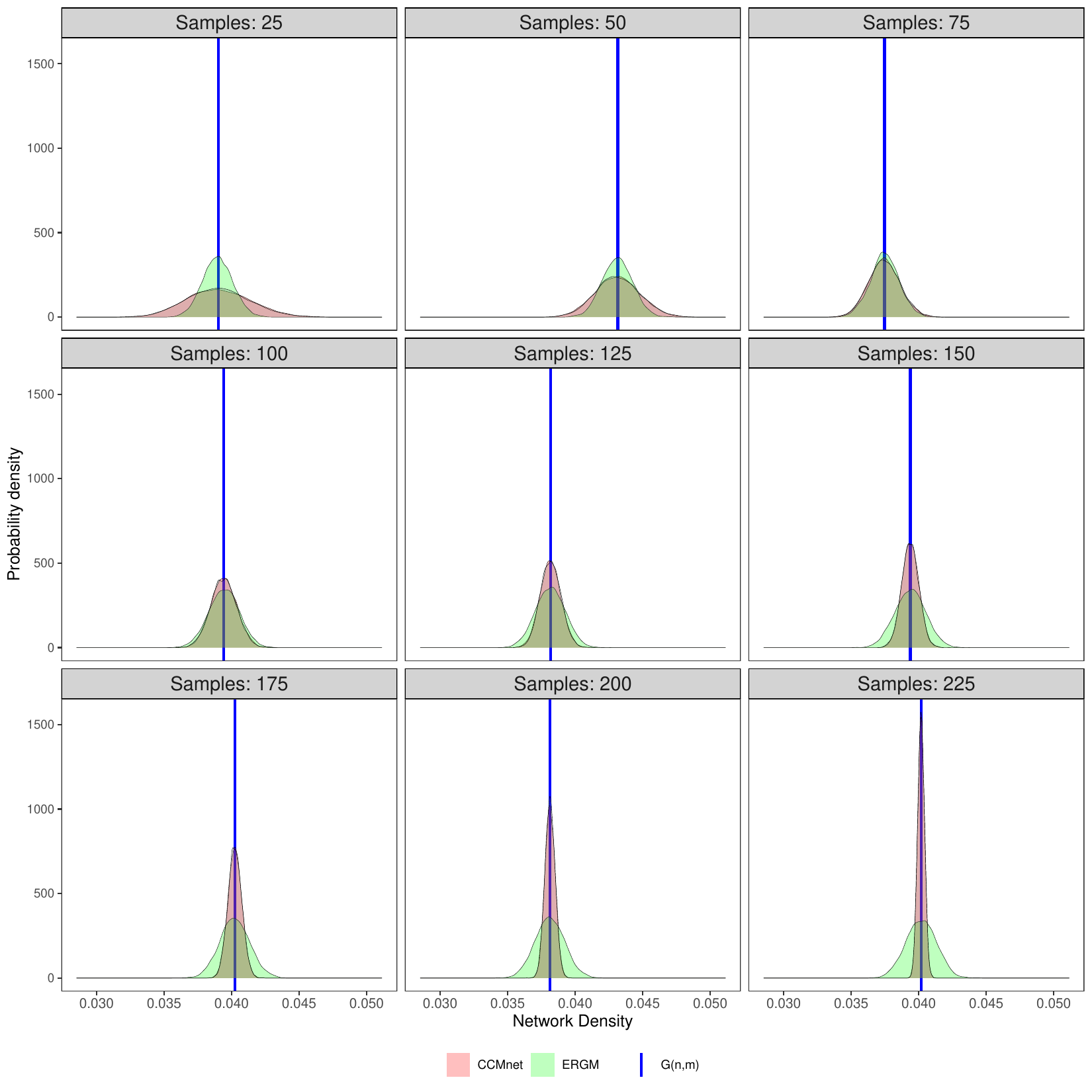}
\caption{Network density distributions for within-network sampling across sample sizes $n \in \{25, 50, \dots, 225\}$. Within each plot, network statistic distributions for the CCM (red), ERGM (green), and $G(n,m)$ model (blue) are presented.}
\label{fig:CCM_ex_density_within_school_beta}
\end{figure}

Figure~\ref{fig:CCM_ex_density_within_school_beta} presents the network density distributions for the CCM (red), ERGM (green), and $G(n,m)$ model (blue) across varying sample sizes. As the sample size increases, the CCM distributions narrow significantly, reflecting the increasing information gained from the larger sample. In contrast, the benchmark models fail to accurately mirror the posterior variance because they lack a mechanism to incorporate the joint influence of the prior distribution and the sampling design on the distribution of the density parameter.

These comparisons highlight that while standard $G(n,m)$ and ERGM specifications are bound to a fixed point estimate (e.g., the posterior mean), \pkg{CCMnet} allows the generated network ensemble to naturally reflect the true level of structural uncertainty induced by the sampling design.

\section{Additional Applications} \label{sec:add_applications}

CCMs have been utilized in a variety of settings where network structure is partially observed, uncertain, or evolving. Prior work has applied the framework to network model selection \citep{goyal2021investigation}, the integration of multiple heterogeneous data sources to estimate network properties \citep{goyal2023estimating}, and the prediction of future structure in dynamic settings \citep{goyal2020dynamic}. By placing probability distributions directly on network properties, \pkg{CCMnet} provides a practical implementation for these tasks, allowing researchers to incorporate information from disparate data sources and propagate uncertainty into predictive network ensembles.

\section{Discussion} \label{sec:discussion}

This paper introduces \pkg{CCMnet}, a software framework for specifying, sampling, and assessing CCMs for networks. CCMs provide a principled method for explicitly modeling structural uncertainty. While traditional approaches often result in ensembles with fixed properties or model-determined levels of variability, \pkg{CCMnet} allows researchers to define a continuous spectrum of uncertainty—ranging from near-exact constraints to broad empirical distributions. Furthermore, this flexibility provides a transparent framework that unifies and extends several classic network modeling approaches.

A central contribution of \pkg{CCMnet} is the translation of this theoretical framework into a practical tool. Implementing CCMs requires evaluating ratios of congruence class sizes to ensure correct MCMC sampling---a task closely connected to graph enumeration problems that are not supported by existing network modeling software. \pkg{CCMnet} implements these calculations, allowing researchers to apply CCMs without coding the combinatorial ratios or MCMC proposal algorithm for each new property specification.

While this paper focuses on edge counts, degree distributions, and mixing patterns, the CCM framework is not limited to these properties. In principle, any network summary for which the graph space can be partitioned into congruence classes may be incorporated. However, extending CCMs to additional network properties requires deriving and implementing the corresponding congruence class size ratios, which remains a mathematically challenging task. As a result, further methodological work on graph enumeration and combinatorial approximations will directly expand the scope of the CCM framework.

In summary, \pkg{CCMnet} provides a practical implementation of Congruence Class Models. These models effectively address the modeling gap between hard and soft constraints by enabling the generation of posterior predictive network ensembles that incorporate uncertainty in empirical estimates.

\section*{Acknowledgments}

This research is supported by grants from the National Institutes of Health (R01 AI-147441, R01 MH-132151, R01 AI-138901, R01 LM-014193, and P30 AI-036214). This project was also made possible by cooperative agreement CDC-RFA-FT-23-0069 (Award: 1 NU38FT000006-01-00) from the CDC’s Center for Forecasting and Outbreak Analytics. Its contents are solely the responsibility of the authors and do not necessarily represent the official views of the Centers for Disease Control and Prevention.
Conflict of Interest: None.

%Bibliography
\bibliography{references}  

\end{document}